\documentclass[%
 reprint,
superscriptaddress,
 amsmath,amssymb,
 aps,
]{revtex4-2}
\usepackage{graphicx}
\usepackage{dcolumn}
\usepackage{color}
\usepackage{comment}
\usepackage{multirow}
\usepackage{hyperref}
\usepackage{float}
\makeatletter
\let\newfloat\newfloat@ltx
\makeatother
\usepackage{algorithm}
\usepackage{algpseudocode}
\usepackage{bm}
\usepackage{mathrsfs}
\usepackage{float}
\usepackage{amsmath}
\usepackage{warpcol}
\graphicspath{{Images/}}

\usepackage{xcolor}
\definecolor{red1}{rgb}{0.6,0,0}

\begin{document}

\title{Genetic-tunneling driven energy optimizer for spin systems}

\newcommand{\KTH}{Department of Applied Physics, School of Engineering Sciences, KTH Royal Institute of Technology, 
AlbaNova University Center, SE-10691 Stockholm, Sweden}

\newcommand{\KTHeecs}{Division of Computational Science and Technology, School of Electrical Engineering and Computer Science, KTH Royal Institute of Technology, AlbaNova University Center, SE-10691 Stockholm, Sweden}

\newcommand{\SeRC}{SeRC (Swedish e-Science Research Center), KTH Royal Institute of Technology, SE-10044 Stockholm, Sweden}
\newcommand{\Uppsala}{Department of Physics and Astronomy, Uppsala University, Box 516, SE-75120 Uppsala, Sweden}
\newcommand{\Orebro}{School of Science and Technology, \"Orebro University, SE-701 82, \"Orebro, Sweden}
\newcommand{\Stockholm}{Department of Materials and Environmental Chemistry, Stockholm University, SE-10691 Stockholm, Sweden}
\newcommand{\UppsalaChem}{Department of Chemistry - Ångström Laboratory, Uppsala University, Box 538, Uppsala, SE-751 21, Sweden}
\newcommand{\CASszbio}{CAS Key Laboratory of Quantitative Engineering Biology, Shenzhen Institute of Synthetic Biology, Shenzhen Institute of Advanced Technology, Chinese Academy of Sciences, Shenzhen 518055, China}

\author{Qichen Xu}
    \affiliation{\KTH}
    \affiliation{\SeRC}
\author{Zhuanglin Shen}
    \affiliation{\CASszbio}
\author{Manuel Pereiro}
    \affiliation{\Uppsala}
    \author{Erik Sjöqvist}
    \affiliation{\Uppsala}
\author{Pawel Herman}
    \affiliation{\KTHeecs}
    \affiliation{\SeRC}
\author{Olle Eriksson}
    \affiliation{\Uppsala}
\author{Anna Delin}
    \affiliation{\KTH}
    \affiliation{\SeRC}

\date{\today}

\begin{abstract}
A long-standing and difficult problem in, e.g., condensed matter physics is how to find the ground state of a complex many-body system where the potential energy surface has a large number of local minima. Spin systems containing complex and/or topological textures, for example spin spirals or magnetic skyrmions, are prime examples of such systems. We propose here a genetic-tunneling-driven variance-controlled optimization approach, and apply it to two-dimensional magnetic skyrmionic systems. The approach combines a local energy-minimizer backend and a metaheuristic global search frontend. The algorithm is naturally concurrent, resulting in short user execution time. We find that the method performs significantly better than simulated annealing (SA). Specifically, we demonstrate that for the Pd/Fe/Ir(111) system, our method correctly and efficiently identifies the experimentally observed spin spiral, skyrmion lattice and ferromagnetic ground states as a function of external magnetic field. To our knowledge, no other optimization method has until now succeeded to do this.
We envision that our findings will pave the way for evolutionary computing in mapping out phase diagrams for spin systems in general.

\end{abstract}

\maketitle

\section{Main}
Optimization algorithms are central in many areas of physics, for instance whenever one is dealing with complex systems such as the many-body problem.
A common challenge is to find the global minimum in the potential energy surface (PES) describing the system. The PES is often extremely complicated with numerous local minima, making it very hard to identify the global minimum. 
Interestingly, these types of systems can often be described using the language of spin models. Therefore, spin models have found wide use in areas not directly connected to magnetism, e.g., percolation theory, protein folding, and stock market trading models~\cite{Cuevas2016}.

In this work, we specifically focus on the important class of two-dimensional (2D) spin systems with complex magnetic interactions -- interactions that give rise to frustration or convoluted magnetic textures, such as for instance skyrmions or other magnetic topological structures~\cite{Muhlbauer2009,Yu2018,PhysRevLett.95.197204}.
For such systems, a number of optimization algorithms have been developed, e.g., gradient-descent based methods, Monte Carlo approaches, and methods based on spin dynamics\cite{speight2020skyrmions,eriksson2017atomistic,muller2019spirit}.
A recurring problem however is the tendency to get trapped into one of the many local energy minima in the PES (the freezing problem) rather than converging toward the global minimum within a reasonable amount of time.

To attempt to improve on this situation, a viable route is to explore meta-heuristic optimization methods, aimed at efficiently exploring the search space in order to find near–optimal solutions.
In particular, Markov chain Monte Carlo (MCMC) based heat-bath methods -- a group of non-gradient sampling algorithms -- have proven themselves to be both effective and robust as regards searching for low-energy states at finite temperature\cite{hastings1970monte,roy2020convergence}.
Unfortunately, in their current implementations\cite{skubic2008method,verlhac2022thermally}, there is still a great deal of prior knowledge needed -- e.g., appropriate initial guesses and manual convergence analysis -- in order to achieve acceptable results.

In order to remedy these shortcomings, several hybrid approaches based on the idea of combining metaheuristic algorithms and typical optimization approaches, e.g., hybrid Monte Carlo\cite{wang2015comparing}, neural annealing optimization\cite{hibat2021variational} or neural evolutionary methods have been proposed\cite{chen2022neural,whitelam2021neuroevolutionary}. However, these approaches have mainly been designed for the Ising model and may not be ideal when handling realistic magnetic materials with long-range interactions.

Here, inspired by the idea of stochastic tunneling \cite{wenzel1999stochastic} hybrid metaheuristic approaches\cite{d2021gga,kapoor2022bayesian}, and evolutionary approaches~\cite{chen2022neural,whitelam2021neuroevolutionary,Zunger2005} a metaheuristic energy minimization approach is proposed and tested for magnetic systems with topological magnetic textures. 
Specifically, it combines an evolutionary algorithm with a local optimizer. The evolutionary algorithm provides a way to select which new regions of the PES to explore, and the local optimizer finds the nearest local minima in the selected regions. We have selected to name our method  "genetic tunneling optimizer" (GTO) to highlight its ability to tunnel energetically inaccessible regions \cite{wenzel1999stochastic} of the PES using not only mutations but also cross-over operators i.e., a "genetic" approach.
We analyze the performance of our proposed algorithm, which is designed to have a high ability to escape from local traps during energy minimization. 
In particular, we investigate the efficiency of our algorithm by performing simulations of a 2D monolayer with model exchange parameters that give rise to Bloch-type skyrmions as well as the experimentally well-studied  Pd/Fe/Ir(111) system, which exhibits a Néel-type skyrmionic phase. 
Our results show that the proposed approach is indeed highly efficient in finding the global minimum in each of these test systems.

\begin{figure*}
    \centering
    \includegraphics[width=16cm]{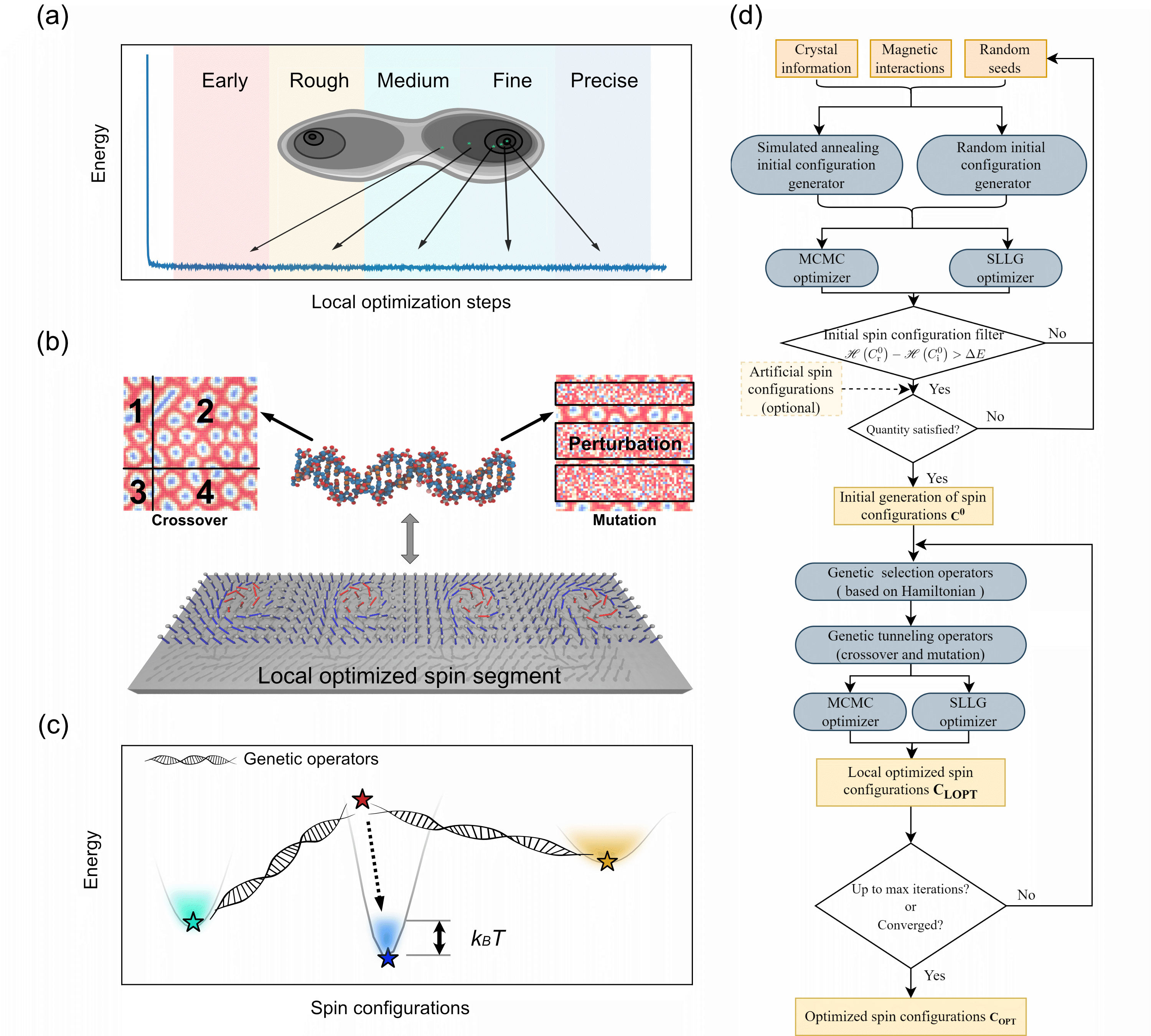}   
    \caption {
    (a) Conceptual illustration of the variance-threshold controlled localized optimization process to find low energy spin configurations. The contour map in the middle shows an example of a PES. The darker zones in the contour map represent lower energy areas. The arrows connect the local optimization process and the configuration point in the PES. At the top of the figure, five colored blocks, i.e., Early, Rough, Medium, Fine, and Precise, denote different converge levels of the local search algorithm. 
    The "Early" level means the lowest convergence, and the "Precise" level represents the highest convergence.
    (b) Conceptual illustration of how the genetic operators are applied to the spin system. The whole process involves three subprocesses, i.e., spin-configuration segmentation, crossover, and perturbation-based mutation (for details, see the Method section). In the spin configuration segmentation part, different from conventional binary-based genetic representation, the configuration-space spin textures are viewed as information carriers conceptually similar to chromosomes in biological systems. The spin textures are divided into several segments that can be used in the same way as gene segments.
    Examples of the square-crossover and mutation operators are shown in the top of Figure (b) (for details about these genetic tunneling operators, see the Method section). (c) Conceptual illustration of how  the genetic tunnelling operators tunnel through the energy barriers and enable a heuristic search for spin configurations with lower energy, with the ultimate aim of finding the global minimum. The curve represents the PES, and the colored stars represent optimized spin configurations corresponding to local minima of the PES at temperature $T$. The height of the shadow-colored region equals $k_BT$. 
    (d) Flowchart of the entire procedure. The dark yellow boxes at the top represent input data that need to be prepared before execution, and the light yellow boxes represent generated spin configurations. The light yellow box with a dashed boundary represents an optional choice. The gray rounded rectangles represent operations. The white diamond-shaped boxes represent conditional statements. 
    Notation in the flow chart: $\mathscr{H}(C^0_r)$ and $\mathscr{H}(C^0_i)$ represent the energy of a generated spin configuration corresponding to a local minimum in the PES and the energy of any spin configuration that has already been selected as part of the initial parent generation, respectively. $\Delta E$ is a threshold energy difference set so that spin configurations that are too similar to the already selected ones are discarded -- see Eq.\ref{eq:threshold_criterion} for more details. $\mathbf{C^0}$ is the initial sets of selected spin configurations, constituting the first parent generation. Finally, $\mathbf{C_{LOPT}}$ and $\mathbf{C_{OPT}}$ are the sets of local and final optimized spin configurations, respectively.}
    \label{fig:idea}
\end{figure*}

\subsection{Spin system parameterization}

Searching for the ground state of a spin system at zero kelvin can be reformulated as finding the global minimum of the PES defined by a many-body Hamiltonian describing the magnetic interactions between the constituents of the system. In the present work, we use a Heisenberg-type classical atomistic spin Hamiltonian of the form
\begin{equation}\label{eq:param-hamiltonian}
\begin{aligned}
\mathscr{H}
&=
-\sum_{i \neq j} J_{i j} \mathbf{S}_{i} \cdot \mathbf{S}_{j}-
\sum_{i \neq j} \mathbf{D}_{i j}\cdot\left(\mathbf{S}_{i} \times \mathbf{S}_{j}\right) \\&- \sum_{i} \mathbf{B}^{\mathrm{ext}} \cdot \mathbf{S}_{i}  - \sum_{i}K^{\mathrm{U}}\left(\mathbf{S}_{i} \cdot \mathbf{e}_{z}\right)^{2},
\end{aligned}
\end{equation}
where $\mathbf{S}_{i}$ is the spin moment at site $i$.  $J_{i j}$, $\mathbf{D}_{i j}$, $K^{\mathrm{U}}$, $\mathbf{e}_{z}$ and $\mathbf{B}^{\mathrm{ext}}$ are Heisenberg exchange interactions, Dzyaloshinskii–Moriya interactions, uniaxial anisotropy, the easy axis vector, and the applied field, respectively. Typically, these four Hamiltonian terms are sufficient for a good description of the system one wishes to analyze (see, e.g., Ref.\cite{eriksson2017atomistic}). Note that for some magnetic systems, additional terms may be relevant to include, e.g., the biquadratic exchange coupling or the four-ring-interaction \cite{eriksson2017atomistic}.

In this work, we test our algorithm by studying two different systems -- an artificial model system, and a system with realistic materials-specific magnetic interaction parameters. In the artificial model system, we use only nearest-neighbor interactions and the parameters are chosen to generate Bloch-type skyrmions. The second system is the Pd/Fe/Ir(111) system. Here, we used more than thirty neighbor-interactions and all parameters were calculated by means of ab-initio density functional theory (DFT)\cite{Miranda2022parameters,bessarab2018}. In the present work, we have considered spin configurations of 100 $\times$ 100 atomic spins, but this number can naturally change, depending on the studied system. All interaction data for both systems can be found in the GitHub repository (See data availability section).
\begin{figure*}[ht]
    \centering
    \includegraphics[width=16cm]{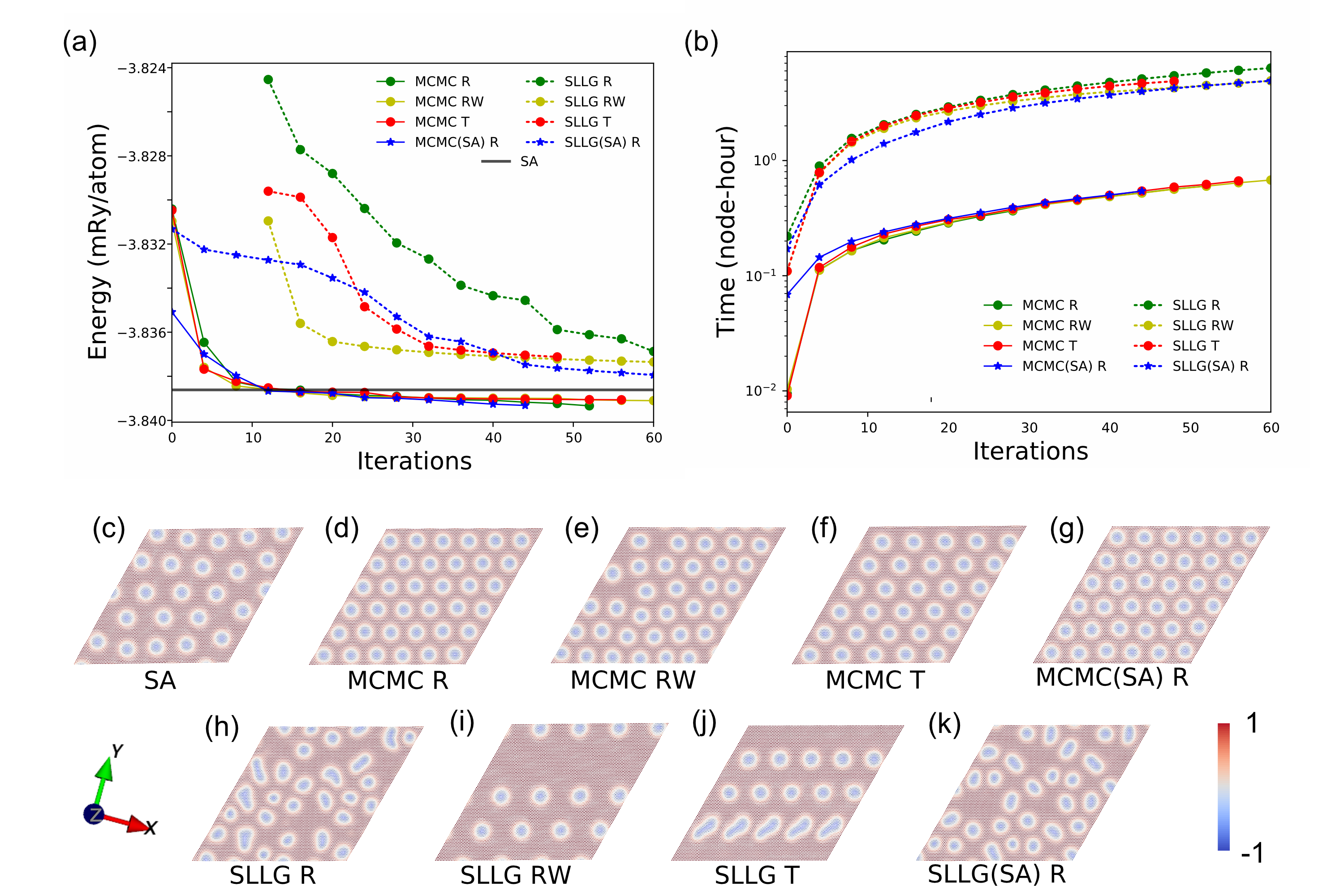}   
     \caption {(a) Performance benchmark of the genetic tunneling optimizer at 0.1\,mK. The simulation system is Pd/Fe/Ir(111) with an external magnetic field of 2.7~T directed out of the plane. Here MCMC-R, MCMC-RW, and MCMC-T represent Markov chain Monte Carlo local optimizers with Rank (R), Roulette Wheel (RW), and Tournament (T) genetic selection, respectively (for details, see the Method section). The  SLLG-R, SLLG-RW, and SLLG-T represent spin-dynamics based optimizers that solve the stochastic Landau–Lifshitz–Gilbert (SLLG) equation with an artificial damping value of 0.4 combined with Rank(R), Roulette Wheel(RW) and Tournament(T) genetic selection, respectively. The abbreviation SA stands for simulated annealing, the results of which is used as reference energy. For better comparison, the results from SLLG R, SLLG RW, and SLLG T are shown from generation 12 of the algorithm (see the Method section). (b) The time consumption in units of node-hours for each optimizer. All of the simulations are performed on an Intel Xeon Gold 6130 CPU node with 32 cores without concurrent processes. (c)-(k) Visualization of the final spin configurations from each optimizer. Here, blue color means that the spin points downwards away from the reader, whereas red color means that the spin points upward, toward the reader. White color indicates that the spin direction is parallel to the plane.
     }
    \label{fig:Result1}
\end{figure*}
\subsection{Outline of the genetic tunneling procedure}
Finding the global minimum in a complex PES of a spin system such as the one described above is commonly a non-deterministic polynomial-time hard (NP-hard) problem~\cite{hibat2021variational}. These problems are very challenging, which has prompted the development of various heuristic methods, for example simulated annealing (SA).
To solve this global energy optimization problem, we propose a genetic tunneling algorithm. The complete details are described in the Method section, here we outline the most salient features. The procedure is illustrated schematically in Figure~\ref{fig:idea}, where, in particular, a flow chart of the method can be found in panel (d).
Our approach builds on genetic algorithms that serve as an optimisation scheme by mimicking the flow of genetic material through generations with an evolutionary tendency for finding better solutions. In our method, segments of spin configurations are conceptually represented as genes, each assigned a quality tag in terms of the underlying energy. In this way, following the logic of genetic algorithm-based optimisation, our method aims at finding the global minimum.
 Importantly, the overall aim of the genetic algorithm is to allow us to reach the global minimum with minimal numerical effort, while we avoid becoming trapped in metastable configurations (i.e., local minima). 

The workflow used here can be summarized into two parts: finding spin configurations corresponding to local mimina in the PES and using these spin configurations to perform "genetic tunneling" over the PES.

The very first step in the procedure is of course to provide basic information about the system to the algorithm -- input which consists of physical information such as crystal lattice, system size, atomic positions, and magnetic interactions. An initial set of preliminary spin configurations is then created. These spin configurations serve as a first coarse guess and are typically far from any energy minimum. 
For each of these preliminary configurations, a local optimization module is invoked, relaxing all magnetic orientations so as to reach, for each of them, the closest local minimum. A selection also takes place, so that the spin configurations that become members of the first parent generation are not too similar to each other. The end result of this step is a diverse set of initial-generation spin configurations representing local minima in the PES.

Then follows the metaheuristic search. Here, the set of initial-generation spin configurations representing local minima are segmented into pieces.
The pieces are then combined together in new ways and also subjected to some "mutation" by adding random noise, see Figure~\ref{fig:idea}(b). This allows us to reach new, unexplored, parts of the PES.
The process is then repeated until convergence, i.e., new local minima are identified, segments of the corresponding spin configurations are combined, and yet new parts of the PES are reached, with the aim of identifying spin configurations with still lower energies than the ones in the previous generations. Additional details on the various parts making up the genetic tunneling method can be found in the Method section.
\begin{figure*}
    \centering
    \includegraphics[width=16cm]{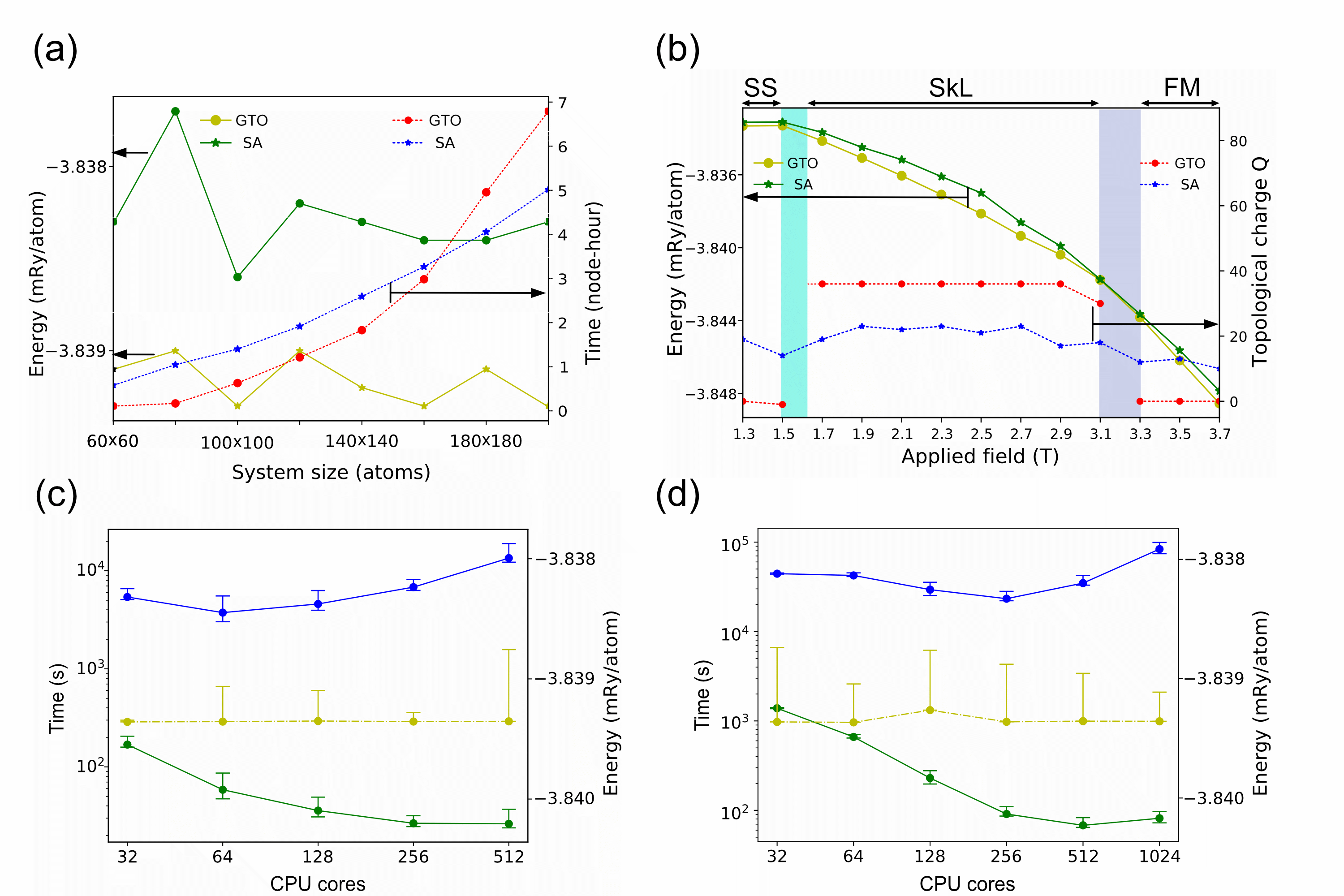}   
    \caption {Performance analysis of the genetic tunneling algorithm at low temperature on the Fe/Pd/Ir (111) system with different system sizes and applied fields. Due to limited space in the figures, we use the label "GTO" to refer to GTO-MCMC-R computations. SA stands for simulated annealing. (a) Predicted ground-state energy and simulation execution time as a function of system size. 
    (b) Total energy and topological charge as a function of applied magnetic field. The size of the simulated system is $100 \times 100$, and the temperature is set to 0.1\,mK. SS, SkL, and FM represent the spin spiral, skyrmion lattice, and ferromagnetic states, respectively.  The cyan- and purple-coloured blocks indicate transition zones between the states.
    (c) First-generation simulation user execution time (green symbols) and core  execution time (blue symbols), and predicted ground-state energy (yellow symbols) as a function of the number of CPU cores for a $100 \times 100$ spin system. The lines are guides for the eye. (d) Same as in (c), but for a $200 \times 200$ spin system.  Each point in (c) and (d) represents 5 simulations. The time error bars in (c) and (d) show the highest, lowest, and average first-generation execution time. The energy error bars in (c) and (d) indicate the highest and lowest predicted ground state energies.}
    \label{fig:Result2}
\end{figure*}
\subsection{Algorithm performance comparisons and optimization}
Each different part of our method can be implemented in various ways. In order to optimize the performance of our method we have therefore systematically combined a number of possible implementations of each part and compared the resulting performance.
Specifically, we combine 
three typical genetic selection operators, i.e., Rank (R), Tournament (T), and Roulette Wheel (RW), with two different local optimizer backends -- the 
MCMC-based optimizer and an optimizer based on the stochastic Landau–Lifshitz–Gilbert (SLLG) equation.
Finally, for the purpose of analyzing the impact of starting from a pre-optimized initial spin configuration, an initial spin configuration generated with SA was tested as initial guess, in addition to simply using a randomly generated initial spin configuration.

We have set the temperature to a very low value, 0.1~mK, in order to obtain highly converged energies and spin configurations of the local minima. This enables a more fair comparison of the performance of the various implementations, since random noise in the optimization process is reduced. For an explanation of the role of temperature in the algorithms and how we use the term "ground state" in relation to that, see the Method section.
The simulations are performed for a spin Hamiltonian describing the two-dimensional Pd/Fe/Ir(111) system.

The results  are presented in Figure~\ref{fig:Result1}.
To facilitate comparison, a classical MCMC-based SA optimization with a fine temperature mesh and $2.5 \times 10^6$ steps is used as baseline. These baseline results are denoted SA in Figure~\ref{fig:Result1}. For more details on the SA setup we have used, see the Method section.
A clear trend in Figure~\ref{fig:Result1}(a) is that all tested combinations indeed manage to find spin configurations with gradually lower energy with each iteration. However, there is a significant difference in performance between the MCMC-based implementations and the spin-dynamics based ones. 
Specifically, we find that with a finite number of local optimization steps and global searching epochs, all spin-dynamics based optimizers may not achieve convergence. 
In general, the implementations with an MCMC-based backend perform significantly better, yielding consistently lower energies for the same number of iterations. 
In fact, all tested genetic tunneling operators combined with the MCMC backend invariably reach convergence and find solutions with lower energy compared to the other tested methods, including the baseline reference energy obtained by SA. 
We also measured and compared the computing time for all implementations, and found that the ones with the MCMC backend were faster on average, see Figure~\ref{fig:Result1}(b).

The Figures~\ref{fig:Result1}(d)-(k) show the spin configurations after 60 iterations for each tested implementation. All implementations with the MCMC backend (Figures~\ref{fig:Result1}(d)-(g)) find a hexagonal skyrmion lattice, with a unit cell size in good agreement with previous studies\cite{dupe2014tailoring,romming2013writing}.
In contrast, none of the implementations based on the spin-dynamics backend (Figures~\ref{fig:Result1}(h)-(k)) manage to identify a stable hexagonal skyrmion lattice within 60 iterations.

All in all, we find that the Rank selection genetic operator algorithm in combination with the MCMC backend appears to be the best choice. Furthermore, we find that starting from a completely random spin configuration is a good choice, since no significant gain could be identified in the tests where we instead started from an initial configuration found using SA. 
Thus, in the remainder of this work, we use the algorithm corresponding to option (d) in Figure~\ref{fig:Result1}, i.e., MCMC with Rank selection and random initial input, and refer to it as GTO-MCMC-R.
\begin{figure*}
    \centering
    \includegraphics[width=16cm]{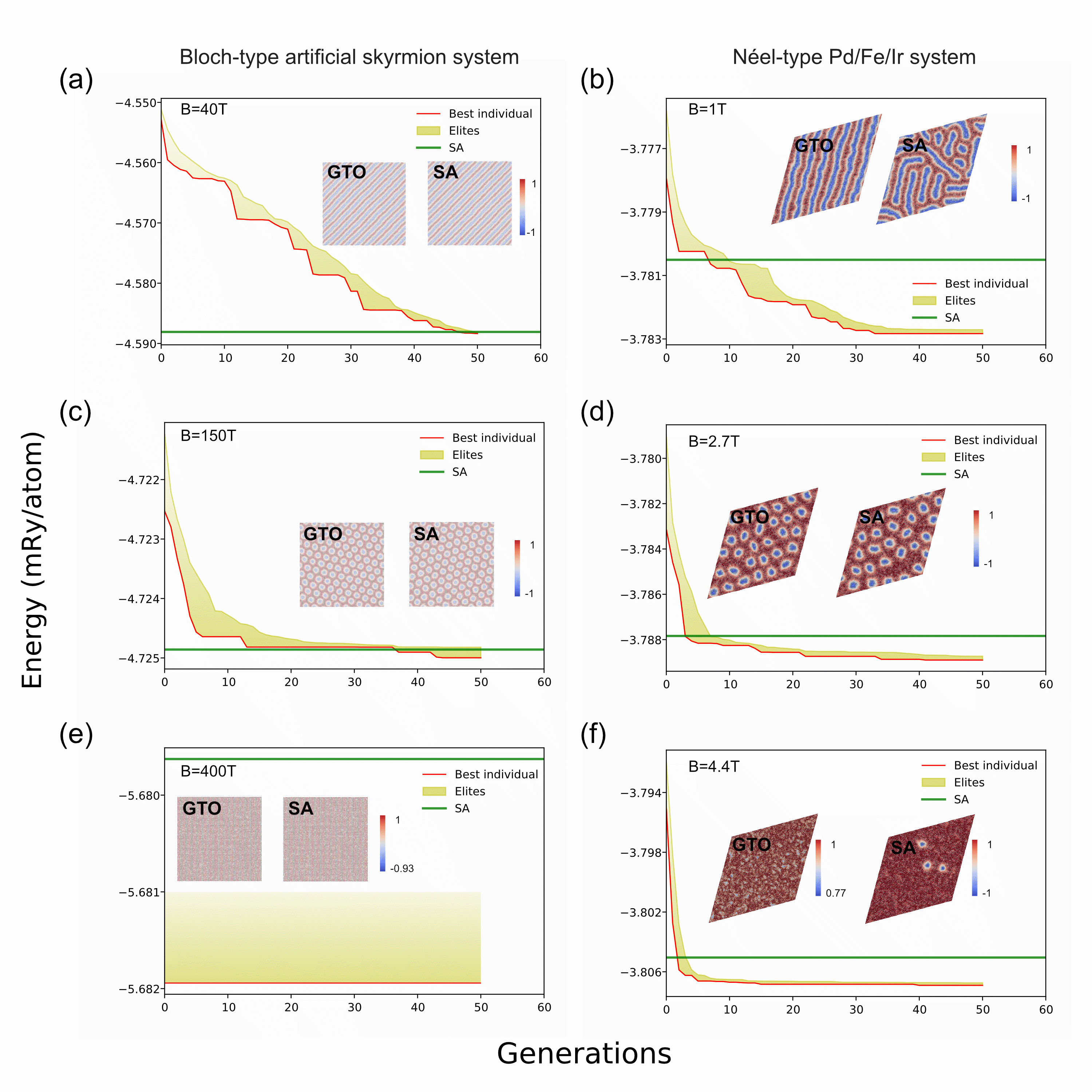}   
    \caption {Searches for the ground state in an artificial Bloch-type skyrmion system and a Néel-type skyrmion system Pd/Fe/Ir(111) that are described in section A, using a simulation temperature of 8~K. 
    Due to limited space in the figures, we use the label "GTO" to refer to GTO-MCMC-R computations. SA stands for simulated annealing.
    Panels (a), (b), and (c) show the simulation results for the artificial Bloch-type skyrmion system  with 40~T, 150~T, and 400~T applied field, respectively. Panels (d), (e), and (f) show the simulation results for the Pd/Fe/Ir(111) system with 1.0~T, 2.7~T, and 4.4~T applied field, respectively. The ground states in (a)-(b), (c)-(d), and (e)-(f) are a spin spiral, a skyrmion lattice, and the ferromagnetic state, respectively.
    In each panel, the red line represents the energy of the best individual of each generation, the yellow band represents the energy distribution of elite individuals~\cite{ahn2003elitism,katoch2021review} in each generation, and the green line represents the energy of the spin configuration predicted using SA. To show the optimization process in detail, we set the convergence limit to an extremely low value. Based on this, all optimizations will run up to the preset maximum iteration threshold of 50 in this section.  The real-space spin configurations found by GTO-MCMC-R and SA  are visualized in the middle of each panel. 
    Here, blue color means that the spin points downwards away from the reader, whereas red color means that the spin points upward, toward the reader. White color indicates that the spin direction is parallel to the plane.
    }
    \label{fig:Result3}
\end{figure*}
{}

\subsection{Effect of system size and applied field on performance}
In Figure~\ref{fig:Result2}, we summarize how our method performs as a function of system size and applied magnetic field, and compare with the corresponding performance of SA. In all these simulations the temperature was set to a low value -- 0.1~mK and the tests were done for the Fe/Pd/Ir(111) system.
Figure~\ref{fig:Result2}(a) shows total energy (left y-scale) and execution time (right y-scale) as a function of system size. Here, the magnetic field was set to 2.7~T and all simulations were performed on a single node with 32 CPU cores. 
We find that GTO-MCMC-R consistently identifies lower energy states compared to SA for all tested system sizes.
GTO-MCMC-R  converges faster than SA for system sizes up to about 160x160 but scales less well than SA for larger system sizes.
We also evaluated the performance of the GTO-MCMC-R  as a function of applied field, see Figure~\ref{fig:Result2}(b). The left y-scale shows the total energy of the system, and the right y-scale the total topological charge (or skyrmion number).  Just as before, we see that GTO-MCMC-R consistently identifies lower energy states compared to SA. 

The most remarkable result here is however how GTO-MCMC-R very successfully can be used to distinctly identify the three different phases of the system and their regions of stability with respect to the magnetic field. This is because the method predicts the topological charge with excellent precision. Below 1.5~T the spin-spiral state is the ground state  and above 3.3~T the ground state is ferromagnetic, both with zero topological charge. From about 1.6~T to around  3~T, a hexagonal skyrmion lattice phase is predicted, with a constant topological charge of 36 (corresponding to a 
$6 \times 6$ skyrmion lattice) for the investigated system size. In contrast, the SA simulations predict nonzero topological charge over the entire investigated magnetic field interval. Moreover, the topological charge computed with SA oscillates visibly in the region where there should be a stable hexagonal skyrmion lattice. In the regions where the topological charge should be zero, SA predicts a topological charge only slightly different compared to the skyrmion-lattice region.

Our GTO-MCMC-R simulations predict that the zone where the skyrmion lattice is the ground state is somewhat wider than what was found in Ref.~\cite{Miranda2022parameters}.
The blue and gray areas indicate transition zones in the GTO-MCMC-R simulations, in which the topological charge changes as the system changes its ground state.  
Energy differences between configurations with various topological charges in the second transition zone can be found in the Supplementary section. 
Finally, in Figures~\ref{fig:Result2}(c) and (d), we show how the GTO-MCMC-R execution time scales with the number of CPU cores used. As already mentioned, our method lends itself well to concurrent computing, which becomes increasingly important the larger the addressed system is. In both figures, the left y-scale is execution time and the right y-scale is the computed total energy of the system. We show user execution time (the dark green circles) as well as core execution time, i.e. the user execution time multiplied with the used number of cores (dark blue circles).
In Figure~\ref{fig:Result2}(c), the size of the tested system is $100 \times 100$, whereas in Figure~\ref{fig:Result2}(d) it is four times larger -- $200 \times 200$. 
As regards the computed total energy per atom (the light green circles), it is found to be effectively constant, as expected.
From the core execution time data, we see that the method scales very well with the number of cores. For the $100 \times 100$ system in Figure~\ref{fig:Result2}(c), the core execution time is smallest for 64 cores, but user execution time can still be gained up to 256 cores. For the $200 \times 200$ system, the corresponding numbers are 256 and 512. 

\begin{figure*}
    \centering
    \includegraphics[width=16cm]{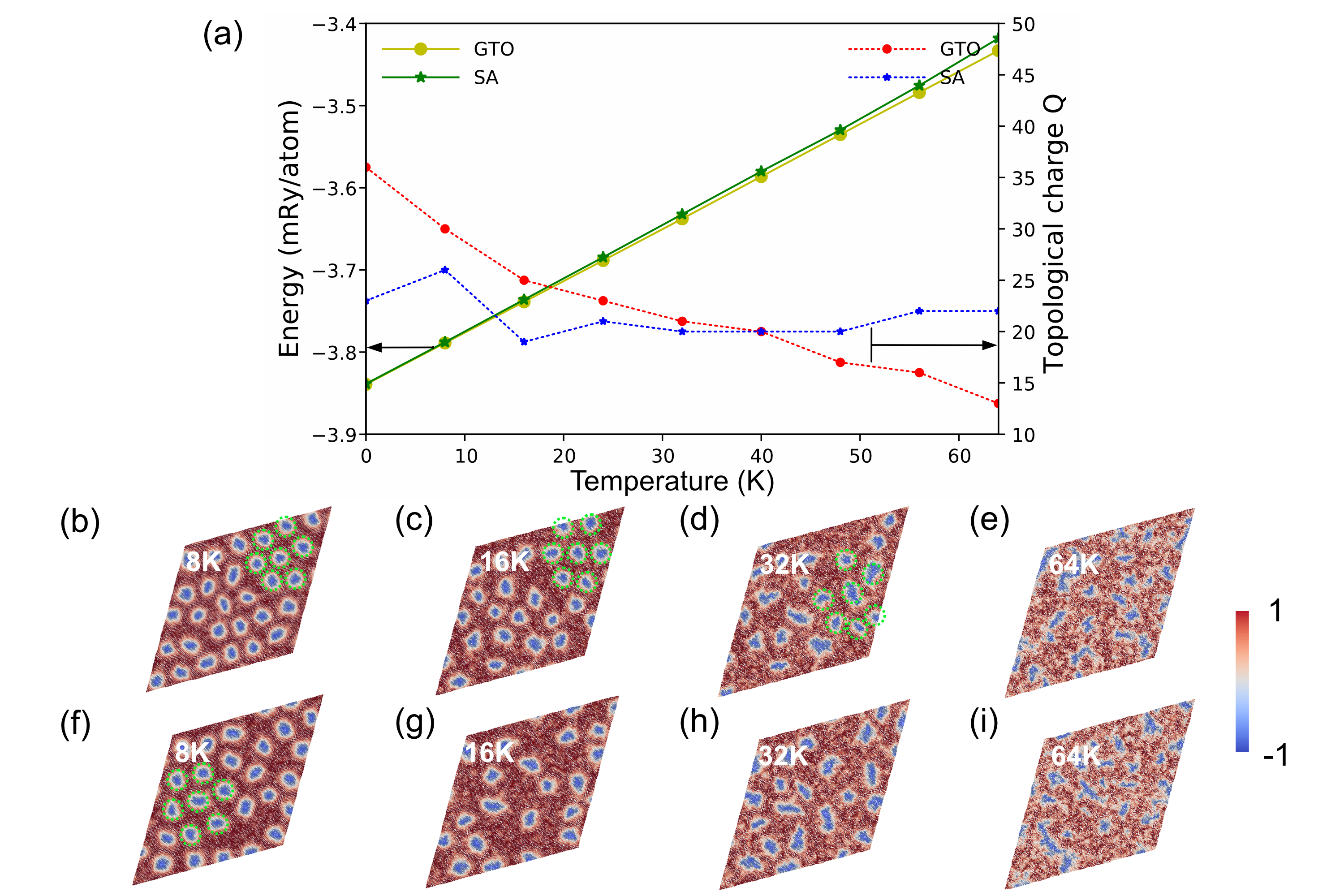}   
    \caption {Ground-state search for the Pd/Fe/Ir (111) system as a function of temperature. The applied magnetic field is set to 2.7~T. 
    Due to limited space in the figures, we use the label "GTO" to refer to GTO-MCMC-R computations. SA stands for simulated annealing.
    (a) The ground state energy (left y-axis) and topological charge (right y-axis).  (b)-(e) Real-space plots of ground-state spin configurations identified using GTO-MCMC-R for four different temperatures. (f)-(i) Real-space plots of ground-state spin configurations identified using SA for four different temperatures.. The green dashed lines point out patterns with hexagonal ordering of skyrmions. The color scheme in the plots (b)-(e) is the same as in Figure~\ref{fig:Result1}(c)-(k).}
    \label{fig:Result4}
\end{figure*}
\subsection{Computed ground state as a function of applied magnetic field with constant temperature }

In this section, we present results from investigations on how the genetic tunneling algorithm performs at a temperature of 8~K for different applied fields. Results for two simulated systems are shown in Figure~\ref{fig:Result3}. The left column (Figure~\ref{fig:Result3} (a/c/e)), shows simulations on an artificial frustrated spin system that exhibits a spin spiral state (a), a Bloch-type skyrmion state (c), and ferromagnetic state (e) at low, medium, and high applied field, respectively. The simulated system in the right column in Figure~\ref{fig:Result3} (b/d/f) is a Pd/Fe/Ir(111) monolayer, which also has three states at different fields, but instead of Bloch skyrmions, it contains Néel-type skyrmions.

As shown in Figure~\ref{fig:Result3} (a/c/e), both the GTO-MCMC-R and SA can find the ground state of the artificial system with different applied fields, but as regards the energy itself, the genetic tunneling algorithm is more successful in finding spin configurations with lower energy. This result is also reflected in the real-space spin configuration visualization, especially in Figure~\ref{fig:Result3} (c). Evidently, while magnetic moments in both systems follow the Boltzmann distribution at 8~K, the hexagonal skyrmion lattice found with the genetic tunneling algorithm is more stable.

From Figure~\ref{fig:Result3} (b/d/e), we find that also for a more complex, real system, the GTO-MCMC-R still performs better than SA. When the applied field is 1~T, the ground state predicted by the genetic tunneling optimizer is a twisty spin spiral state which is in remarkably good agreement with the experimental data in \cite{romming2013writing}. However, SA gives a hybrid phase that contains both spin spirals and bubbles. At 2.7~T, GTO-MCMC-R finds a more reasonable hexagonal skyrmion lattice compared to SA. Finally, at 4.4~T, the GTO-MCMC-R correctly finds the ferromagnetic state (with some fluctuations due to finite temperature) whereas in the SA solution there are several complex structures (bubbles or skyrmions). To summarize, the final system energy that we obtain from the genetic tunneling optimizer is lower than what we obtain using SA in both systems addressed. Thus, we conclude that our algorithm works well for a range of magnetic fields in the presence of thermal fluctuations.

\subsection{Computed ground state as a function of temperature with constant applied magnetic field}

Next, we analyzed the performance of our algorithm as a function of temperature. In all these simulations the applied magnetic field was set to 2.7~T (i.e., in the middle of the interval where the skyrmion lattice is the ground state) and the tests were done for the Fe/Pd/Ir(111) system. The results are shown in Figure~\ref{fig:Result4}. 
In this case, we find that the ground state energies predicted by GTO-MCMC-R (yellow symbols) and SA (green symbols) are broadly similar. However, as the temperature increases, the difference increases, with GTO-MCMC-R consistently finding slightly lower energy states, see Figure~\ref{fig:Result4}(a). We also computed the topological charge of the predicted ground states. In the GTO-MCMC-R simulations, the topological charge decreases with temperature, which is consistent with experimental results\cite{romming2013writing,PhysRevB.101.214445} and also what one would expect. In contrast, SA does not reproduce this trend. Additional data in the form of spin configurations are shown in Figure~\ref{fig:Result4} (b)-(i), where the top row are the GTO-MCMC-R spin configuration results and the bottom row the corresponding SA results. In the GTO-MCMC-R images, remnants of the ground-state hexagonal skyrmion crystal pattern can be seen up to 32~K, whereas such patterns are absent in the SA results above 8~K. It is reasonable to expect that the zero kelvin hexagonal skyrmion lattice breaks up gradually with increasing temperature, and therefore we believe that these images further strengthen the conclusion that GTO-MCMC-R has superior performance compared to SA.

\section{Discussion and Conclusions}
To the best of our knowledge, this is the first study establishing a genetic-tunneling optimization protocol for complex spin systems with long-range interactions. 
The algorithm presented in this work has general applicability to spin systems, and is here shown to successfully find the ground state for monolayer spin systems at finite temperatures and at a variety of applied magnetic fields. The approach contains two essential parts: (1) a variance-threshold controlled local optimizer, which includes a MCMC optimizer and a spin-dynamic optimizer and (2) a spin-configuration space genetic tunneling metaheuristic search module. 
The algorithm is designed to be able to escape from a local minimum through the use of genetic tunneling operators and find the global minimum for a given system without any initial guess. 
The efficiency of our genetic tunneling protocol is investigated on both a simple artificial system with magnetic frustration and a Pd/Fe/Ir(111) monolayer that includes complex Heisenberg and Dzyaloshinskii–Moriya interactions, as calculated from DFT. The results indicate that the GTO-MCMC-R has better performance than SA when it comes to finding stable spin configurations as a function of external parameters like temperature and applied magnetic field. Most noteworthy, we here considered the spiral structure, the skyrmion lattice, and the ferromagnetic state. Our method successfully finds all three ground states as a function of magnetic field, and also correctly identifies the transition regions in between. To our knowledge, no other theoretical method has yet demonstrated such an ability. 
It can also be concluded that the performance of the algorithm is limited neither by the system size, geometry, nature of the magnetic interactions, temperature, nor the applied field strength.

In practice, for optimal performance of the protocol, the various hyperparameters will of course need to be fine-tuned depending on the system under consideration. For example, a non-optimal variance threshold may increase the risk of premature convergence. 

In conclusion, we have explored a genetic tunneling protocol, which is designed to predict the magnetic ground state of a classical spin Hamiltonian at finite temperature. We demonstrate that our method is robust for two-dimensional systems, both for a simpler model systems and for the more complex Pd/Fe/Ir(111) system. We envision that our findings will pave the way for evolutionary computing in finding the ground state of magnetic systems, e.g., for magnets with non-trivial topology and spin glass systems. Since complex systems in a very general sense can be cast into the language of spin Hamiltonians, it also appears possible that the here suggested protocols will find applications in other areas of solid-state science, or even in fields outside the natural sciences.

\section{Method}
A metaheuristic algorithm can interactively guide and modify the operations of subordinate heuristics, to efficiently produce preferable solutions within a high-dimensional search space\cite{beheshti2013review,abdel2018metaheuristic}. 
Representative metaheuristic algorithms include Simulated Annealing (SA), Particle Swarm Optimization (PSO)\cite{bautu2007particle} and Genetic Algorithms (GA)\cite{oda2005search}. It has been shown that a hybrid algorithm that combines the algorithms mentioned above with a local optimizer, e.g., gradient descent, can be an efficient way to solve the global optimization problem within a complex configuration space\cite{ting2015hybrid}.

In this study we introduce a genetic tunneling strategy in the form of a hybrid algorithm  -- connecting the local minimization approach shown in Figure~\ref{fig:idea}(a) 
to a metaheuristic genetic-tunneling module -- with the aim of finding the global energy minimum of the spin system. For an overview of the procedure, see the flow chart in Figure~\ref{fig:idea}(d). In the following sections, we describe the main implemented algorithms and how they interact. In the current code, there are a number of additional optional functionalities implemented, which are not discussed here since they are not of central importance for the overall performance, e.g., the Elite selection protocol\cite{ahn2003elitism,katoch2021review}.

\subsection{Selection and genetic-tunneling operators} 

Below, we describe the evolutionary parts of our hybrid algorithm.
Each generation consists of $N_p$ spin configurations, i.e., for the $k^{\mathrm{th}}$ generation we have
\begin{equation}
\begin{aligned}
\mathbf{C}^k=\left\{C^k_i \mid i=1,2, \cdots, N_p\right\},
\end{aligned}
\end{equation}
where $\mathbf{C}^k$ is the set of all spin configurations in generation $k$, and $C^k_i$ is one spin configuration. Starting from the $k^{\mathrm{th}}$ generation of $N_p$ spin configurations, each corresponding to a local minimum in the PES, generation $k+1$ is produced in the following way. The first step is to select spin configurations in generation $k$ that will be used as parents for a new spin configuration in generation $k+1$. We have implemented three different selection operators -- Roulette Wheel (RW), Rank (R), and Tournament (T). 
With the Roulette Wheel selection operator, the probability of choosing a candidate spin configuration $C^k_i$ to become a member of the breeding pool is equal to
\begin{equation}
\begin{aligned}
P({C^k_i})=\frac{\mathscr{H}_i}{\sum_{i=1}^{N_{p}} \mathscr{H}_i} ,
\end{aligned}
\label{probable}
\end{equation}
where $\mathscr{H}_i$ is the Hamiltonian (i.e., the energy) of spin configuration $C^k_i$. 
The Rank and Tournament selection operators are described as algorithms in the Supplementary section. In the present implementation, $N_p$ is set to 64 and we select a total of four spin configurations out of these to breed one new spin configuration. 

The next step is to create the new spin configuration from the selected parents. 
To perform this step, we invoke three tunneling operators -- square crossover, linear crossover and mutation. The pseudocode for the quared-crossover and linear-crossover operators can be found in the Supplementary section.
In the square-crossover scheme, we first generate two random split ratios, and use them to split each parent spin configurations along the $x$- and $y$-directions into four rectangles, see Figure~\ref{fig:idea}(b). Then, we combine these generated parts using a randomized procedure to create a new spin configuration with the same dimensions as the spin configurations in the parent generation. Finally, the mutation operator is added. This operator consists of adding a Gaussian random noise to all spins in the new configuration, using the Rodrigues rotation formula\cite{bessarab2015method}.
Thus, the square-crossover scheme consists of two operators -- the square-crossover operator and the mutation operator.
The linear-crossover scheme is very similar to the square-crossover scheme, with the difference that we instead split each parent spin configuration along only one direction. These steps are then repeated until we have $N_p$ new spin configurations.
As a side note, we mention that the square-crossover and linear-crossover tunneling operators we introduce here are especially well suited for systems with clearly contained spin textures, such as skyrmions and spin spirals.

Before the new generation of spin configurations can be used as parents, a local energy optimizer is invoked, to guarantee that all spin configurations in the new generation correspond to local minima in the PES. The local energy optimizers we have used for this step are described next.

\subsection{Variance-controlled local energy optimizer at finite temperature}
A local optimization of a spin configuration involves finding a local minimum starting from a given initial guess. 
We have implemented two different types of local optimizers: an
MCMC-based optimizer and one based on SLLG. These two methods have been proven to be robust and efficient in describing complex spin systems\cite{furuya2015semi,eriksson2017atomistic}. We describe both in some detail below.

The MCMC optimizer is of Metropolis type and performs energy minimization under finite temperature by using the transition probability $P_t$ between two spin configurations in the Markov chain:
\begin{equation}
\begin{aligned}
P_t= \begin{cases}\exp \left(-\frac{\Delta E}{k_BT}\right), & \text { if } \Delta E>0 \\ 1, & \text { otherwise }\end{cases}
\end{aligned}
\end{equation}
where $\Delta E$, $k_B$, and $T$ are the energy difference between spin configurations, the Boltzmann constant, and the temperature of the system. For a given initial spin configuration, the method will iteratively minimize the energy of the  system.

The second approach we use to optimize spin configurations is the spin-dynamics based optimizer, which uses the SLLG equation to simulate the time evolution of atomic magnetic moments. This method is able to reach a spin configuration near a local energy minimum from a given initial state, since -- when Gilbert damping\cite{furuya2015semi} is included in the simulations -- energy is allowed to dissipate from the system. The atomistic SLLG equation reads
\begin{equation}
\begin{aligned}
\frac{d \mathbf{m}_i}{d t}=&-\gamma_{\mathbf{L}} \mathbf{m}_i \times\left(\mathbf{B}_i+\mathbf{B}_i^{\mathrm{f}}\right)\\
&-\gamma_{\mathrm{L}} \frac{\alpha}{m_i} \mathbf{m}_i \times\left[\mathbf{m}_i \times\left(\mathbf{B}_i+\mathbf{B}_i^{\mathrm{f}}\right)\right] ,
\end{aligned}
\end{equation}
where $\mathbf{B}_i$ the effective magnetic field at site $i$ and $\mathbf{B}_i^{\mathrm{f}}$ is the stochastic magnetic field corresponding to the thermal fluctuations present in a heat bath with temperature $T$. In this equation, the first term represents the precessional motion of the atomic magnetic moments, while the second term describes the damping motion. In the expression above, $\gamma_{\mathrm{L}}$ is the renormalized gyromagnetic ratio, which is calculated from
\begin{equation}
\begin{aligned}
\gamma_{\mathrm{L}}=\frac{\gamma}{\left(1+\alpha^2\right)} ,
\end{aligned}
\end{equation}
where $\gamma$ is the gyromagnetic ratio and $\alpha$ is the isotropic Gilbert damping constant.
In the present work, we use both the Metropolis MCMC and the SLLG local optimization schemes as implemented in the Uppsala Atomistic Spin Dynamics (UppASD) package\cite{skubic2008method}. 

In order to automatically stop the minimization process when a desirable convergence level has been reached, we use a variance threshold $\operatorname{Var}(\mathscr{H})$. For the spin Hamiltonian used here, this threshold is defined as
\begin{equation}
\begin{aligned}
\operatorname{Var}(\mathscr{H})=\frac{1}{n} \sum_{i=1}^{n} \left(\mathscr{H}_i-\left\langle \mathscr{H}\right\rangle\right)^2 ,
\end{aligned}
\end{equation}
where $\left\langle \mathscr{H} \right\rangle$ stands for the expectation value of the spin Hamiltonian, and the sum is over the $n$ last iteration steps. In the present work, we typically set $n$ between 10 and 100.

We end this subsection with a note on nomenclature: as is evident from the above, in our simulation method the temperature must always be set to a finite value, it cannot be zero. Therefore, in this work, the term "ground state" should be taken to mean the lowest lying minimum at the set simulation temperature, see Figure~\ref{fig:idea}(c). 

\subsection{Initialization and termination}
{\it Initialization} The algorithm starts from an initial parent generation $\mathbf{C}^0$ of spin configurations (generation zero).
We have implemented two ways of producing these initial configurations. The first method starts by simply producing random spin configurations using random numbers together with the crystal information and magnetic interactions defining the studied spin system. The generated spin configurations are subsequently relaxed to the closest lying local minimum, using one of the local energy optimizers described above.We select $N_p$ spin configurations by using the criterion that any two configurations must not be too close in energy, i.e., 
\begin{equation}
\label{eq:threshold_criterion}
\mathscr{H}\left({C}^{0}_{r}\right) - \mathscr{H}\left({C}^{0}_{i
}\right))
>\Delta E .
\end{equation}
Here, $\Delta E$ is a threshold energy difference, that guarantees that configurations $C^0_r$ and $C^0_i$ are not too similar. In this work, we use values ranging from $10^{-4}$ to $10^{-6}$ mRy/atom for $\Delta E$. 
In practice, this implies that the number of random spin configurations generated will typically be much larger than $N_p$, in order to be able to select $N_p$ spin configurations fulfilling the selection criterion above. The $N_p$ selected spin configurations constitute $\mathbf{C}^0$.
The second method is based on simulated annealing (SA). Here, we use a relatively coarse temperature mesh so that the SA-based initialization does not become too time consuming. 
Just as in the first method, a large number of spin configurations are generated, a local minimizer is invoked to find the corresponding local minima, and
$N_p$ spin configurations are selected using the criterion in Eq.~\ref{eq:threshold_criterion}.
Specifically, in the present work, the SA simulations were performed using a temperature mesh of only four points (the simulated temperature, and then 20~K, 50~K, and 200~K added to the simulated temperature). At each temperature, we performed 2000 Metropolis steps. 

{\it Termination}
There are two stop criteria set for the search process in this work -- a maximum number of allowed iterations (i.e., number of produced generations) and a convergence criterion.
In this work we have set the maximum number of iterations to around 60. However, our optimization algorithm often finds a converged solution (according to the convergence criterion explained below) within a significantly smaller number of iterations. 
Our convergence criterion is designed in the following way. For each generation, the variance over the set of spin configurations is computed. Thus, for each generation, we compute
\begin{equation}
\label{eq:variance2}
\begin{aligned}
\operatorname{Var}(\mathscr{H})=\frac{1}{N_p} \sum_{i=1}^{N_p} \left(\mathscr{H}_i-\left\langle \mathscr{H}\right\rangle\right)^2 ,
\end{aligned}
\end{equation}
where $\left\langle \mathscr{H} \right\rangle$ stands for the expectation value of the spin Hamiltonians over all spin configurations in the set.
As the optimization proceeds, the spin configurations within each generation will become more and more similar to each other, as they approach the ground state. Therefore, when the variance in Eq.~\ref{eq:variance2} decreases below a predefined threshold, the procedure is deemed to have converged, and is stopped. At this point, the final spin configuration or configurations $\mathbf{C_{OPT}}$ representing the best solution can be extracted (see the flowchart in Figure~\ref{fig:idea}(d)).

\section{Data availability}
All data needed for reproducing the results can be found in the GitHub repository \url{https://github.com/MXJK851/GTO-2D}

\section{Code availability}
All code of GTO-2D is available at \url{https://github.com/MXJK851/GTO-2D} under the GPL-3.0 license.

\begin{acknowledgments}
The authors acknowledge financial support from the Knut and Alice Wallenberg Foundation 
through Grant No. 2018.0060. Q.X. acknowledges the China Scholarship Council (201906920083), O.E. acknowledges eSSENCE, the Swedish Research Council (VR) and the ERC (project FASTCORR-Grant No. 854843). 
A.D. acknowledges financial support from
the Swedish Research Council (VR) through Grants
No. 2019-05304 and 2016-05980.
The computations/data handling were enabled by
resources provided by the Swedish National Infrastructure for Computing (SNIC), partially funded
by the Swedish Research Council through grant agreement No. 2016-07213.
The authors would also like to extend our gratitude to Fredrik Dehlin, Liang Chen, Ivan Miranda, Filipp N. Rybakov, Heike Herper and Liuzhen Yang for their valuable comments and discussions during the writing of our paper.
\end{acknowledgments}

\bibliography{apssamp}

\begin{thebibliography}{33}%
\makeatletter
\providecommand \@ifxundefined [1]{%
 \@ifx{#1\undefined}
}%
\providecommand \@ifnum [1]{%
 \ifnum #1\expandafter \@firstoftwo
 \else \expandafter \@secondoftwo
 \fi
}%
\providecommand \@ifx [1]{%
 \ifx #1\expandafter \@firstoftwo
 \else \expandafter \@secondoftwo
 \fi
}%
\providecommand \natexlab [1]{#1}%
\providecommand \enquote  [1]{``#1''}%
\providecommand \bibnamefont  [1]{#1}%
\providecommand \bibfnamefont [1]{#1}%
\providecommand \citenamefont [1]{#1}%
\providecommand \href@noop [0]{\@secondoftwo}%
\providecommand \href [0]{\begingroup \@sanitize@url \@href}%
\providecommand \@href[1]{\@@startlink{#1}\@@href}%
\providecommand \@@href[1]{\endgroup#1\@@endlink}%
\providecommand \@sanitize@url [0]{\catcode `\\12\catcode `\$12\catcode
  `\&12\catcode `\#12\catcode `\^12\catcode `\_12\catcode `\%12\relax}%
\providecommand \@@startlink[1]{}%
\providecommand \@@endlink[0]{}%
\providecommand \url  [0]{\begingroup\@sanitize@url \@url }%
\providecommand \@url [1]{\endgroup\@href {#1}{\urlprefix }}%
\providecommand \urlprefix  [0]{URL }%
\providecommand \Eprint [0]{\href }%
\providecommand \doibase [0]{http://dx.doi.org/}%
\providecommand \selectlanguage [0]{\@gobble}%
\providecommand \bibinfo  [0]{\@secondoftwo}%
\providecommand \bibfield  [0]{\@secondoftwo}%
\providecommand \translation [1]{[#1]}%
\providecommand \BibitemOpen [0]{}%
\providecommand \bibitemStop [0]{}%
\providecommand \bibitemNoStop [0]{.\EOS\space}%
\providecommand \EOS [0]{\spacefactor3000\relax}%
\providecommand \BibitemShut  [1]{\csname bibitem#1\endcsname}%
\let\auto@bib@innerbib\@empty
\bibitem [{\citenamefont {De~las Cuevas}\ and\ \citenamefont
  {Cubitt}(2016)}]{Cuevas2016}%
  \BibitemOpen
  \bibfield  {author} {\bibinfo {author} {\bibfnamefont {G.}~\bibnamefont
  {De~las Cuevas}}\ and\ \bibinfo {author} {\bibfnamefont {T.~S.}\ \bibnamefont
  {Cubitt}},\ }\href {\doibase 10.1126/science.aab3326} {\bibfield  {journal}
  {\bibinfo  {journal} {Science}\ }\textbf {\bibinfo {volume} {351}},\ \bibinfo
  {pages} {1180–1183} (\bibinfo {year} {2016})}\BibitemShut {NoStop}%
\bibitem [{\citenamefont {Mühlbauer}\ \emph {et~al.}(2009)\citenamefont
  {Mühlbauer}, \citenamefont {Binz}, \citenamefont {Jonietz}, \citenamefont
  {Pfleiderer}, \citenamefont {Rosch}, \citenamefont {Neubauer}, \citenamefont
  {Georgii},\ and\ \citenamefont {Böni}}]{Muhlbauer2009}%
  \BibitemOpen
  \bibfield  {author} {\bibinfo {author} {\bibfnamefont {S.}~\bibnamefont
  {Mühlbauer}}, \bibinfo {author} {\bibfnamefont {B.}~\bibnamefont {Binz}},
  \bibinfo {author} {\bibfnamefont {F.}~\bibnamefont {Jonietz}}, \bibinfo
  {author} {\bibfnamefont {C.}~\bibnamefont {Pfleiderer}}, \bibinfo {author}
  {\bibfnamefont {A.}~\bibnamefont {Rosch}}, \bibinfo {author} {\bibfnamefont
  {A.}~\bibnamefont {Neubauer}}, \bibinfo {author} {\bibfnamefont
  {R.}~\bibnamefont {Georgii}}, \ and\ \bibinfo {author} {\bibfnamefont
  {P.}~\bibnamefont {Böni}},\ }\href {\doibase 10.1126/science.1166767}
  {\bibfield  {journal} {\bibinfo  {journal} {Science}\ }\textbf {\bibinfo
  {volume} {323}},\ \bibinfo {pages} {915–919} (\bibinfo {year}
  {2009})}\BibitemShut {NoStop}%
\bibitem [{\citenamefont {Yu}\ \emph {et~al.}(2018)\citenamefont {Yu},
  \citenamefont {Koshibae}, \citenamefont {Tokunaga}, \citenamefont {Shibata},
  \citenamefont {Taguchi}, \citenamefont {Nagaosa},\ and\ \citenamefont
  {Tokura}}]{Yu2018}%
  \BibitemOpen
  \bibfield  {author} {\bibinfo {author} {\bibfnamefont {X.~Z.}\ \bibnamefont
  {Yu}}, \bibinfo {author} {\bibfnamefont {W.}~\bibnamefont {Koshibae}},
  \bibinfo {author} {\bibfnamefont {Y.}~\bibnamefont {Tokunaga}}, \bibinfo
  {author} {\bibfnamefont {K.}~\bibnamefont {Shibata}}, \bibinfo {author}
  {\bibfnamefont {Y.}~\bibnamefont {Taguchi}}, \bibinfo {author} {\bibfnamefont
  {N.}~\bibnamefont {Nagaosa}}, \ and\ \bibinfo {author} {\bibfnamefont
  {Y.}~\bibnamefont {Tokura}},\ }\href {\doibase 10.1038/s41586-018-0745-3}
  {\bibfield  {journal} {\bibinfo  {journal} {Nature}\ }\textbf {\bibinfo
  {volume} {564}},\ \bibinfo {pages} {95–98} (\bibinfo {year}
  {2018})}\BibitemShut {NoStop}%
\bibitem [{\citenamefont {Tchernyshyov}\ and\ \citenamefont
  {Chern}(2005)}]{PhysRevLett.95.197204}%
  \BibitemOpen
  \bibfield  {author} {\bibinfo {author} {\bibfnamefont {O.}~\bibnamefont
  {Tchernyshyov}}\ and\ \bibinfo {author} {\bibfnamefont {G.-W.}\ \bibnamefont
  {Chern}},\ }\href {\doibase 10.1103/PhysRevLett.95.197204} {\bibfield
  {journal} {\bibinfo  {journal} {Phys. Rev. Lett.}\ }\textbf {\bibinfo
  {volume} {95}},\ \bibinfo {pages} {197204} (\bibinfo {year}
  {2005})}\BibitemShut {NoStop}%
\bibitem [{\citenamefont {Speight}\ and\ \citenamefont
  {Winyard}(2020)}]{speight2020skyrmions}%
  \BibitemOpen
  \bibfield  {author} {\bibinfo {author} {\bibfnamefont {M.}~\bibnamefont
  {Speight}}\ and\ \bibinfo {author} {\bibfnamefont {T.}~\bibnamefont
  {Winyard}},\ }\href@noop {} {\bibfield  {journal} {\bibinfo  {journal}
  {Physical Review B}\ }\textbf {\bibinfo {volume} {101}},\ \bibinfo {pages}
  {134420} (\bibinfo {year} {2020})}\BibitemShut {NoStop}%
\bibitem [{\citenamefont {Eriksson}\ \emph {et~al.}(2017)\citenamefont
  {Eriksson}, \citenamefont {Bergman}, \citenamefont {Bergqvist},\ and\
  \citenamefont {Hellsvik}}]{eriksson2017atomistic}%
  \BibitemOpen
  \bibfield  {author} {\bibinfo {author} {\bibfnamefont {O.}~\bibnamefont
  {Eriksson}}, \bibinfo {author} {\bibfnamefont {A.}~\bibnamefont {Bergman}},
  \bibinfo {author} {\bibfnamefont {L.}~\bibnamefont {Bergqvist}}, \ and\
  \bibinfo {author} {\bibfnamefont {J.}~\bibnamefont {Hellsvik}},\ }\href@noop
  {} {\emph {\bibinfo {title} {Atomistic spin dynamics: foundations and
  applications}}}\ (\bibinfo  {publisher} {Oxford university press},\ \bibinfo
  {year} {2017})\BibitemShut {NoStop}%
\bibitem [{\citenamefont {M{\"u}ller}\ \emph {et~al.}(2019)\citenamefont
  {M{\"u}ller}, \citenamefont {Hoffmann}, \citenamefont {Di{\ss}elkamp},
  \citenamefont {Sch{\"u}rhoff}, \citenamefont {Mavros}, \citenamefont
  {Sallermann}, \citenamefont {Kiselev}, \citenamefont {J{\'o}nsson},\ and\
  \citenamefont {Bl{\"u}gel}}]{muller2019spirit}%
  \BibitemOpen
  \bibfield  {author} {\bibinfo {author} {\bibfnamefont {G.~P.}\ \bibnamefont
  {M{\"u}ller}}, \bibinfo {author} {\bibfnamefont {M.}~\bibnamefont
  {Hoffmann}}, \bibinfo {author} {\bibfnamefont {C.}~\bibnamefont
  {Di{\ss}elkamp}}, \bibinfo {author} {\bibfnamefont {D.}~\bibnamefont
  {Sch{\"u}rhoff}}, \bibinfo {author} {\bibfnamefont {S.}~\bibnamefont
  {Mavros}}, \bibinfo {author} {\bibfnamefont {M.}~\bibnamefont {Sallermann}},
  \bibinfo {author} {\bibfnamefont {N.~S.}\ \bibnamefont {Kiselev}}, \bibinfo
  {author} {\bibfnamefont {H.}~\bibnamefont {J{\'o}nsson}}, \ and\ \bibinfo
  {author} {\bibfnamefont {S.}~\bibnamefont {Bl{\"u}gel}},\ }\href@noop {}
  {\bibfield  {journal} {\bibinfo  {journal} {Physical Review B}\ }\textbf
  {\bibinfo {volume} {99}},\ \bibinfo {pages} {224414} (\bibinfo {year}
  {2019})}\BibitemShut {NoStop}%
\bibitem [{\citenamefont {Hastings}(1970)}]{hastings1970monte}%
  \BibitemOpen
  \bibfield  {author} {\bibinfo {author} {\bibfnamefont {W.~K.}\ \bibnamefont
  {Hastings}},\ }\href@noop {} {\emph {\bibinfo {title} {Monte Carlo sampling
  methods using Markov chains and their applications}}}\ (\bibinfo  {publisher}
  {Oxford University Press},\ \bibinfo {year} {1970})\BibitemShut {NoStop}%
\bibitem [{\citenamefont {Roy}(2020)}]{roy2020convergence}%
  \BibitemOpen
  \bibfield  {author} {\bibinfo {author} {\bibfnamefont {V.}~\bibnamefont
  {Roy}},\ }\href@noop {} {\bibfield  {journal} {\bibinfo  {journal} {Annual
  Review of Statistics and Its Application}\ }\textbf {\bibinfo {volume} {7}},\
  \bibinfo {pages} {387} (\bibinfo {year} {2020})}\BibitemShut {NoStop}%
\bibitem [{\citenamefont {Skubic}\ \emph {et~al.}(2008)\citenamefont {Skubic},
  \citenamefont {Hellsvik}, \citenamefont {Nordstr{\"o}m},\ and\ \citenamefont
  {Eriksson}}]{skubic2008method}%
  \BibitemOpen
  \bibfield  {author} {\bibinfo {author} {\bibfnamefont {B.}~\bibnamefont
  {Skubic}}, \bibinfo {author} {\bibfnamefont {J.}~\bibnamefont {Hellsvik}},
  \bibinfo {author} {\bibfnamefont {L.}~\bibnamefont {Nordstr{\"o}m}}, \ and\
  \bibinfo {author} {\bibfnamefont {O.}~\bibnamefont {Eriksson}},\ }\href@noop
  {} {\bibfield  {journal} {\bibinfo  {journal} {Journal of physics: condensed
  matter}\ }\textbf {\bibinfo {volume} {20}},\ \bibinfo {pages} {315203}
  (\bibinfo {year} {2008})}\BibitemShut {NoStop}%
\bibitem [{\citenamefont {Verlhac}\ \emph {et~al.}(2022)\citenamefont
  {Verlhac}, \citenamefont {Niggli}, \citenamefont {Bergman}, \citenamefont
  {Kamber}, \citenamefont {Bagrov}, \citenamefont {Iu{\c{s}}an}, \citenamefont
  {Nordstr{\"o}m}, \citenamefont {Katsnelson}, \citenamefont {Wegner},
  \citenamefont {Eriksson} \emph {et~al.}}]{verlhac2022thermally}%
  \BibitemOpen
  \bibfield  {author} {\bibinfo {author} {\bibfnamefont {B.}~\bibnamefont
  {Verlhac}}, \bibinfo {author} {\bibfnamefont {L.}~\bibnamefont {Niggli}},
  \bibinfo {author} {\bibfnamefont {A.}~\bibnamefont {Bergman}}, \bibinfo
  {author} {\bibfnamefont {U.}~\bibnamefont {Kamber}}, \bibinfo {author}
  {\bibfnamefont {A.}~\bibnamefont {Bagrov}}, \bibinfo {author} {\bibfnamefont
  {D.}~\bibnamefont {Iu{\c{s}}an}}, \bibinfo {author} {\bibfnamefont
  {L.}~\bibnamefont {Nordstr{\"o}m}}, \bibinfo {author} {\bibfnamefont {M.~I.}\
  \bibnamefont {Katsnelson}}, \bibinfo {author} {\bibfnamefont
  {D.}~\bibnamefont {Wegner}}, \bibinfo {author} {\bibfnamefont
  {O.}~\bibnamefont {Eriksson}},  \emph {et~al.},\ }\href@noop {} {\bibfield
  {journal} {\bibinfo  {journal} {Nature Physics}\ }\textbf {\bibinfo {volume}
  {18}},\ \bibinfo {pages} {905} (\bibinfo {year} {2022})}\BibitemShut
  {NoStop}%
\bibitem [{\citenamefont {Wang}\ \emph {et~al.}(2015)\citenamefont {Wang},
  \citenamefont {Machta},\ and\ \citenamefont
  {Katzgraber}}]{wang2015comparing}%
  \BibitemOpen
  \bibfield  {author} {\bibinfo {author} {\bibfnamefont {W.}~\bibnamefont
  {Wang}}, \bibinfo {author} {\bibfnamefont {J.}~\bibnamefont {Machta}}, \ and\
  \bibinfo {author} {\bibfnamefont {H.~G.}\ \bibnamefont {Katzgraber}},\
  }\href@noop {} {\bibfield  {journal} {\bibinfo  {journal} {Physical Review
  E}\ }\textbf {\bibinfo {volume} {92}},\ \bibinfo {pages} {013303} (\bibinfo
  {year} {2015})}\BibitemShut {NoStop}%
\bibitem [{\citenamefont {Hibat-Allah}\ \emph {et~al.}(2021)\citenamefont
  {Hibat-Allah}, \citenamefont {Inack}, \citenamefont {Wiersema}, \citenamefont
  {Melko},\ and\ \citenamefont {Carrasquilla}}]{hibat2021variational}%
  \BibitemOpen
  \bibfield  {author} {\bibinfo {author} {\bibfnamefont {M.}~\bibnamefont
  {Hibat-Allah}}, \bibinfo {author} {\bibfnamefont {E.~M.}\ \bibnamefont
  {Inack}}, \bibinfo {author} {\bibfnamefont {R.}~\bibnamefont {Wiersema}},
  \bibinfo {author} {\bibfnamefont {R.~G.}\ \bibnamefont {Melko}}, \ and\
  \bibinfo {author} {\bibfnamefont {J.}~\bibnamefont {Carrasquilla}},\
  }\href@noop {} {\bibfield  {journal} {\bibinfo  {journal} {Nature Machine
  Intelligence}\ }\textbf {\bibinfo {volume} {3}},\ \bibinfo {pages} {952}
  (\bibinfo {year} {2021})}\BibitemShut {NoStop}%
\bibitem [{\citenamefont {Chen}\ \emph {et~al.}(2022)\citenamefont {Chen},
  \citenamefont {Choo}, \citenamefont {Astrakhantsev},\ and\ \citenamefont
  {Neupert}}]{chen2022neural}%
  \BibitemOpen
  \bibfield  {author} {\bibinfo {author} {\bibfnamefont {A.}~\bibnamefont
  {Chen}}, \bibinfo {author} {\bibfnamefont {K.}~\bibnamefont {Choo}}, \bibinfo
  {author} {\bibfnamefont {N.}~\bibnamefont {Astrakhantsev}}, \ and\ \bibinfo
  {author} {\bibfnamefont {T.}~\bibnamefont {Neupert}},\ }\href@noop {}
  {\bibfield  {journal} {\bibinfo  {journal} {Physical Review Research}\
  }\textbf {\bibinfo {volume} {4}},\ \bibinfo {pages} {L022026} (\bibinfo
  {year} {2022})}\BibitemShut {NoStop}%
\bibitem [{\citenamefont {Whitelam}\ and\ \citenamefont
  {Tamblyn}(2021)}]{whitelam2021neuroevolutionary}%
  \BibitemOpen
  \bibfield  {author} {\bibinfo {author} {\bibfnamefont {S.}~\bibnamefont
  {Whitelam}}\ and\ \bibinfo {author} {\bibfnamefont {I.}~\bibnamefont
  {Tamblyn}},\ }\href@noop {} {\bibfield  {journal} {\bibinfo  {journal}
  {Physical review letters}\ }\textbf {\bibinfo {volume} {127}},\ \bibinfo
  {pages} {018003} (\bibinfo {year} {2021})}\BibitemShut {NoStop}%
\bibitem [{\citenamefont {Wenzel}\ and\ \citenamefont
  {Hamacher}(1999)}]{wenzel1999stochastic}%
  \BibitemOpen
  \bibfield  {author} {\bibinfo {author} {\bibfnamefont {W.}~\bibnamefont
  {Wenzel}}\ and\ \bibinfo {author} {\bibfnamefont {K.}~\bibnamefont
  {Hamacher}},\ }\href@noop {} {\bibfield  {journal} {\bibinfo  {journal}
  {Physical Review Letters}\ }\textbf {\bibinfo {volume} {82}},\ \bibinfo
  {pages} {3003} (\bibinfo {year} {1999})}\BibitemShut {NoStop}%
\bibitem [{\citenamefont {D’Angelo}\ and\ \citenamefont
  {Palmieri}(2021)}]{d2021gga}%
  \BibitemOpen
  \bibfield  {author} {\bibinfo {author} {\bibfnamefont {G.}~\bibnamefont
  {D’Angelo}}\ and\ \bibinfo {author} {\bibfnamefont {F.}~\bibnamefont
  {Palmieri}},\ }\href@noop {} {\bibfield  {journal} {\bibinfo  {journal}
  {Information Sciences}\ }\textbf {\bibinfo {volume} {547}},\ \bibinfo {pages}
  {136} (\bibinfo {year} {2021})}\BibitemShut {NoStop}%
\bibitem [{\citenamefont {Kapoor}\ \emph {et~al.}(2022)\citenamefont {Kapoor},
  \citenamefont {Nukala},\ and\ \citenamefont {Chandra}}]{kapoor2022bayesian}%
  \BibitemOpen
  \bibfield  {author} {\bibinfo {author} {\bibfnamefont {A.}~\bibnamefont
  {Kapoor}}, \bibinfo {author} {\bibfnamefont {E.}~\bibnamefont {Nukala}}, \
  and\ \bibinfo {author} {\bibfnamefont {R.}~\bibnamefont {Chandra}},\
  }\href@noop {} {\bibfield  {journal} {\bibinfo  {journal} {Applied Soft
  Computing}\ }\textbf {\bibinfo {volume} {129}},\ \bibinfo {pages} {109528}
  (\bibinfo {year} {2022})}\BibitemShut {NoStop}%
\bibitem [{\citenamefont {Hart}\ \emph {et~al.}(2005)\citenamefont {Hart},
  \citenamefont {Blum}, \citenamefont {Walorski},\ and\ \citenamefont
  {Zunger}}]{Zunger2005}%
  \BibitemOpen
  \bibfield  {author} {\bibinfo {author} {\bibfnamefont {G.~L.}\ \bibnamefont
  {Hart}}, \bibinfo {author} {\bibfnamefont {V.}~\bibnamefont {Blum}}, \bibinfo
  {author} {\bibfnamefont {M.~J.}\ \bibnamefont {Walorski}}, \ and\ \bibinfo
  {author} {\bibfnamefont {A.}~\bibnamefont {Zunger}},\ }\href@noop {}
  {\bibfield  {journal} {\bibinfo  {journal} {Nature materials}\ }\textbf
  {\bibinfo {volume} {4}},\ \bibinfo {pages} {391} (\bibinfo {year}
  {2005})}\BibitemShut {NoStop}%
\bibitem [{\citenamefont {Miranda}\ \emph {et~al.}(2022)\citenamefont
  {Miranda}, \citenamefont {Klautau}, \citenamefont {Bergman},\ and\
  \citenamefont {Petrilli}}]{Miranda2022parameters}%
  \BibitemOpen
  \bibfield  {author} {\bibinfo {author} {\bibfnamefont {I.~P.}\ \bibnamefont
  {Miranda}}, \bibinfo {author} {\bibfnamefont {A.~B.}\ \bibnamefont
  {Klautau}}, \bibinfo {author} {\bibfnamefont {A.}~\bibnamefont {Bergman}}, \
  and\ \bibinfo {author} {\bibfnamefont {H.~M.}\ \bibnamefont {Petrilli}},\
  }\href {\doibase 10.1103/PhysRevB.105.224413} {\bibfield  {journal} {\bibinfo
   {journal} {Phys. Rev. B}\ }\textbf {\bibinfo {volume} {105}},\ \bibinfo
  {pages} {224413} (\bibinfo {year} {2022})}\BibitemShut {NoStop}%
\bibitem [{\citenamefont {Bessarab}\ \emph {et~al.}(2018)\citenamefont
  {Bessarab}, \citenamefont {Müller}, \citenamefont {Lobanov}, \citenamefont
  {Rybakov}, \citenamefont {Kiselev}, \citenamefont {Jónsson}, \citenamefont
  {Uzdin}, \citenamefont {Blügel}, \citenamefont {Bergqvist}, \citenamefont
  {Delin},\ and\ \citenamefont {et~al.}}]{bessarab2018}%
  \BibitemOpen
  \bibfield  {author} {\bibinfo {author} {\bibfnamefont {P.~F.}\ \bibnamefont
  {Bessarab}}, \bibinfo {author} {\bibfnamefont {G.~P.}\ \bibnamefont
  {Müller}}, \bibinfo {author} {\bibfnamefont {I.~S.}\ \bibnamefont
  {Lobanov}}, \bibinfo {author} {\bibfnamefont {F.~N.}\ \bibnamefont
  {Rybakov}}, \bibinfo {author} {\bibfnamefont {N.~S.}\ \bibnamefont
  {Kiselev}}, \bibinfo {author} {\bibfnamefont {H.}~\bibnamefont {Jónsson}},
  \bibinfo {author} {\bibfnamefont {V.~M.}\ \bibnamefont {Uzdin}}, \bibinfo
  {author} {\bibfnamefont {S.}~\bibnamefont {Blügel}}, \bibinfo {author}
  {\bibfnamefont {L.}~\bibnamefont {Bergqvist}}, \bibinfo {author}
  {\bibfnamefont {A.}~\bibnamefont {Delin}}, \ and\ \bibinfo {author}
  {\bibnamefont {et~al.}},\ }\href {\doibase 10.1038/s41598-018-21623-3}
  {\bibfield  {journal} {\bibinfo  {journal} {Scientific Reports}\ }\textbf
  {\bibinfo {volume} {8}} (\bibinfo {year} {2018}),\
  10.1038/s41598-018-21623-3}\BibitemShut {NoStop}%
\bibitem [{\citenamefont {Dup{\'e}}\ \emph {et~al.}(2014)\citenamefont
  {Dup{\'e}}, \citenamefont {Hoffmann}, \citenamefont {Paillard},\ and\
  \citenamefont {Heinze}}]{dupe2014tailoring}%
  \BibitemOpen
  \bibfield  {author} {\bibinfo {author} {\bibfnamefont {B.}~\bibnamefont
  {Dup{\'e}}}, \bibinfo {author} {\bibfnamefont {M.}~\bibnamefont {Hoffmann}},
  \bibinfo {author} {\bibfnamefont {C.}~\bibnamefont {Paillard}}, \ and\
  \bibinfo {author} {\bibfnamefont {S.}~\bibnamefont {Heinze}},\ }\href@noop {}
  {\bibfield  {journal} {\bibinfo  {journal} {Nature communications}\ }\textbf
  {\bibinfo {volume} {5}},\ \bibinfo {pages} {1} (\bibinfo {year}
  {2014})}\BibitemShut {NoStop}%
\bibitem [{\citenamefont {Romming}\ \emph {et~al.}(2013)\citenamefont
  {Romming}, \citenamefont {Hanneken}, \citenamefont {Menzel}, \citenamefont
  {Bickel}, \citenamefont {Wolter}, \citenamefont {von Bergmann}, \citenamefont
  {Kubetzka},\ and\ \citenamefont {Wiesendanger}}]{romming2013writing}%
  \BibitemOpen
  \bibfield  {author} {\bibinfo {author} {\bibfnamefont {N.}~\bibnamefont
  {Romming}}, \bibinfo {author} {\bibfnamefont {C.}~\bibnamefont {Hanneken}},
  \bibinfo {author} {\bibfnamefont {M.}~\bibnamefont {Menzel}}, \bibinfo
  {author} {\bibfnamefont {J.~E.}\ \bibnamefont {Bickel}}, \bibinfo {author}
  {\bibfnamefont {B.}~\bibnamefont {Wolter}}, \bibinfo {author} {\bibfnamefont
  {K.}~\bibnamefont {von Bergmann}}, \bibinfo {author} {\bibfnamefont
  {A.}~\bibnamefont {Kubetzka}}, \ and\ \bibinfo {author} {\bibfnamefont
  {R.}~\bibnamefont {Wiesendanger}},\ }\href@noop {} {\bibfield  {journal}
  {\bibinfo  {journal} {Science}\ }\textbf {\bibinfo {volume} {341}},\ \bibinfo
  {pages} {636} (\bibinfo {year} {2013})}\BibitemShut {NoStop}%
\bibitem [{\citenamefont {Ahn}\ and\ \citenamefont
  {Ramakrishna}(2003)}]{ahn2003elitism}%
  \BibitemOpen
  \bibfield  {author} {\bibinfo {author} {\bibfnamefont {C.~W.}\ \bibnamefont
  {Ahn}}\ and\ \bibinfo {author} {\bibfnamefont {R.~S.}\ \bibnamefont
  {Ramakrishna}},\ }\href@noop {} {\bibfield  {journal} {\bibinfo  {journal}
  {IEEE Transactions on Evolutionary Computation}\ }\textbf {\bibinfo {volume}
  {7}},\ \bibinfo {pages} {367} (\bibinfo {year} {2003})}\BibitemShut {NoStop}%
\bibitem [{\citenamefont {Katoch}\ \emph {et~al.}(2021)\citenamefont {Katoch},
  \citenamefont {Chauhan},\ and\ \citenamefont {Kumar}}]{katoch2021review}%
  \BibitemOpen
  \bibfield  {author} {\bibinfo {author} {\bibfnamefont {S.}~\bibnamefont
  {Katoch}}, \bibinfo {author} {\bibfnamefont {S.~S.}\ \bibnamefont {Chauhan}},
  \ and\ \bibinfo {author} {\bibfnamefont {V.}~\bibnamefont {Kumar}},\
  }\href@noop {} {\bibfield  {journal} {\bibinfo  {journal} {Multimedia Tools
  and Applications}\ }\textbf {\bibinfo {volume} {80}},\ \bibinfo {pages}
  {8091} (\bibinfo {year} {2021})}\BibitemShut {NoStop}%
\bibitem [{\citenamefont {Lindner}\ \emph {et~al.}(2020)\citenamefont
  {Lindner}, \citenamefont {Bargsten}, \citenamefont {Kovarik}, \citenamefont
  {Friedlein}, \citenamefont {Harm}, \citenamefont {Krause},\ and\
  \citenamefont {Wiesendanger}}]{PhysRevB.101.214445}%
  \BibitemOpen
  \bibfield  {author} {\bibinfo {author} {\bibfnamefont {P.}~\bibnamefont
  {Lindner}}, \bibinfo {author} {\bibfnamefont {L.}~\bibnamefont {Bargsten}},
  \bibinfo {author} {\bibfnamefont {S.}~\bibnamefont {Kovarik}}, \bibinfo
  {author} {\bibfnamefont {J.}~\bibnamefont {Friedlein}}, \bibinfo {author}
  {\bibfnamefont {J.}~\bibnamefont {Harm}}, \bibinfo {author} {\bibfnamefont
  {S.}~\bibnamefont {Krause}}, \ and\ \bibinfo {author} {\bibfnamefont
  {R.}~\bibnamefont {Wiesendanger}},\ }\href {\doibase
  10.1103/PhysRevB.101.214445} {\bibfield  {journal} {\bibinfo  {journal}
  {Phys. Rev. B}\ }\textbf {\bibinfo {volume} {101}},\ \bibinfo {pages}
  {214445} (\bibinfo {year} {2020})}\BibitemShut {NoStop}%
\bibitem [{\citenamefont {Beheshti}\ and\ \citenamefont
  {Shamsuddin}(2013)}]{beheshti2013review}%
  \BibitemOpen
  \bibfield  {author} {\bibinfo {author} {\bibfnamefont {Z.}~\bibnamefont
  {Beheshti}}\ and\ \bibinfo {author} {\bibfnamefont {S.~M.~H.}\ \bibnamefont
  {Shamsuddin}},\ }\href@noop {} {\bibfield  {journal} {\bibinfo  {journal}
  {Int. J. Adv. Soft Comput. Appl}\ }\textbf {\bibinfo {volume} {5}},\ \bibinfo
  {pages} {1} (\bibinfo {year} {2013})}\BibitemShut {NoStop}%
\bibitem [{\citenamefont {Abdel-Basset}\ \emph {et~al.}(2018)\citenamefont
  {Abdel-Basset}, \citenamefont {Abdel-Fatah},\ and\ \citenamefont
  {Sangaiah}}]{abdel2018metaheuristic}%
  \BibitemOpen
  \bibfield  {author} {\bibinfo {author} {\bibfnamefont {M.}~\bibnamefont
  {Abdel-Basset}}, \bibinfo {author} {\bibfnamefont {L.}~\bibnamefont
  {Abdel-Fatah}}, \ and\ \bibinfo {author} {\bibfnamefont {A.~K.}\ \bibnamefont
  {Sangaiah}},\ }\href@noop {} {\bibfield  {journal} {\bibinfo  {journal}
  {Computational intelligence for multimedia big data on the cloud with
  engineering applications}\ ,\ \bibinfo {pages} {185}} (\bibinfo {year}
  {2018})}\BibitemShut {NoStop}%
\bibitem [{\citenamefont {Bautu}\ \emph {et~al.}(2007)\citenamefont {Bautu},
  \citenamefont {Bautu},\ and\ \citenamefont {Luchian}}]{bautu2007particle}%
  \BibitemOpen
  \bibfield  {author} {\bibinfo {author} {\bibfnamefont {A.}~\bibnamefont
  {Bautu}}, \bibinfo {author} {\bibfnamefont {E.}~\bibnamefont {Bautu}}, \ and\
  \bibinfo {author} {\bibfnamefont {H.}~\bibnamefont {Luchian}},\ }in\
  \href@noop {} {\emph {\bibinfo {booktitle} {Ninth International Symposium on
  Symbolic and Numeric Algorithms for Scientific Computing (SYNASC 2007)}}}\
  (\bibinfo {organization} {IEEE},\ \bibinfo {year} {2007})\ pp.\ \bibinfo
  {pages} {415--418}\BibitemShut {NoStop}%
\bibitem [{\citenamefont {Oda}\ \emph {et~al.}(2005)\citenamefont {Oda},
  \citenamefont {Nagao}, \citenamefont {Kitagawa}, \citenamefont {Shigeta},
  \citenamefont {Shoji}, \citenamefont {Nitta}, \citenamefont {Okumura},\ and\
  \citenamefont {Yamaguchi}}]{oda2005search}%
  \BibitemOpen
  \bibfield  {author} {\bibinfo {author} {\bibfnamefont {A.}~\bibnamefont
  {Oda}}, \bibinfo {author} {\bibfnamefont {H.}~\bibnamefont {Nagao}}, \bibinfo
  {author} {\bibfnamefont {Y.}~\bibnamefont {Kitagawa}}, \bibinfo {author}
  {\bibfnamefont {Y.}~\bibnamefont {Shigeta}}, \bibinfo {author} {\bibfnamefont
  {M.}~\bibnamefont {Shoji}}, \bibinfo {author} {\bibfnamefont
  {H.}~\bibnamefont {Nitta}}, \bibinfo {author} {\bibfnamefont
  {M.}~\bibnamefont {Okumura}}, \ and\ \bibinfo {author} {\bibfnamefont
  {K.}~\bibnamefont {Yamaguchi}},\ }\href@noop {} {\bibfield  {journal}
  {\bibinfo  {journal} {International journal of quantum chemistry}\ }\textbf
  {\bibinfo {volume} {105}},\ \bibinfo {pages} {645} (\bibinfo {year}
  {2005})}\BibitemShut {NoStop}%
\bibitem [{\citenamefont {Ting}\ \emph {et~al.}(2015)\citenamefont {Ting},
  \citenamefont {Yang}, \citenamefont {Cheng},\ and\ \citenamefont
  {Huang}}]{ting2015hybrid}%
  \BibitemOpen
  \bibfield  {author} {\bibinfo {author} {\bibfnamefont {T.}~\bibnamefont
  {Ting}}, \bibinfo {author} {\bibfnamefont {X.-S.}\ \bibnamefont {Yang}},
  \bibinfo {author} {\bibfnamefont {S.}~\bibnamefont {Cheng}}, \ and\ \bibinfo
  {author} {\bibfnamefont {K.}~\bibnamefont {Huang}},\ }\href@noop {}
  {\bibfield  {journal} {\bibinfo  {journal} {Recent advances in swarm
  intelligence and evolutionary computation}\ ,\ \bibinfo {pages} {71}}
  (\bibinfo {year} {2015})}\BibitemShut {NoStop}%
\bibitem [{\citenamefont {Bessarab}\ \emph {et~al.}(2015)\citenamefont
  {Bessarab}, \citenamefont {Uzdin},\ and\ \citenamefont
  {J{\'o}nsson}}]{bessarab2015method}%
  \BibitemOpen
  \bibfield  {author} {\bibinfo {author} {\bibfnamefont {P.~F.}\ \bibnamefont
  {Bessarab}}, \bibinfo {author} {\bibfnamefont {V.~M.}\ \bibnamefont {Uzdin}},
  \ and\ \bibinfo {author} {\bibfnamefont {H.}~\bibnamefont {J{\'o}nsson}},\
  }\href@noop {} {\bibfield  {journal} {\bibinfo  {journal} {Computer Physics
  Communications}\ }\textbf {\bibinfo {volume} {196}},\ \bibinfo {pages} {335}
  (\bibinfo {year} {2015})}\BibitemShut {NoStop}%
\bibitem [{\citenamefont {Furuya}\ \emph {et~al.}(2015)\citenamefont {Furuya},
  \citenamefont {Fujisaki}, \citenamefont {Shimizu}, \citenamefont {Uehara},
  \citenamefont {Ataka}, \citenamefont {Tanaka},\ and\ \citenamefont
  {Oshima}}]{furuya2015semi}%
  \BibitemOpen
  \bibfield  {author} {\bibinfo {author} {\bibfnamefont {A.}~\bibnamefont
  {Furuya}}, \bibinfo {author} {\bibfnamefont {J.}~\bibnamefont {Fujisaki}},
  \bibinfo {author} {\bibfnamefont {K.}~\bibnamefont {Shimizu}}, \bibinfo
  {author} {\bibfnamefont {Y.}~\bibnamefont {Uehara}}, \bibinfo {author}
  {\bibfnamefont {T.}~\bibnamefont {Ataka}}, \bibinfo {author} {\bibfnamefont
  {T.}~\bibnamefont {Tanaka}}, \ and\ \bibinfo {author} {\bibfnamefont
  {H.}~\bibnamefont {Oshima}},\ }\href@noop {} {\bibfield  {journal} {\bibinfo
  {journal} {IEEE Transactions on Magnetics}\ }\textbf {\bibinfo {volume}
  {51}},\ \bibinfo {pages} {1} (\bibinfo {year} {2015})}\BibitemShut {NoStop}%
\end{thebibliography}%

\section{Supplementary}
\subsection{Implementation details of selection and genetic tunneling operators}
The pseudocode corresponding to the tournament selection, rank selection, square crossover, and linear crossover algorithms used in this work are provided in algorithm \ref{alg1}, \ref{alg2}, \ref{alg3}, and \ref{alg4}, respectively.

\subsection{Relative energies of different phases in the Pd/Fe/Ir (111) system at 0.1~mK in the transition zone from the skyrmion lattice phase to the ferromagnetic phase}

In Table~\ref{tab:AverageRelativeEnergy}, we list the calculated average relative energies per atom for four different phases in the Pd/Fe/Ir (111) system: the ferromagnetic phase (FM), the isolated skyrmion phase (Sk) and the $6 \times 5 $ and $6 \times 6$ skyrmion lattice phase (SkL). The applied magnetic field is varied from 3.1 to 3.4 T. Zero energy denotes the found ground state in each case. This table proves that the algorithm we propose in this paper indeed correctly identifies the transition state in Figure~\ref{fig:Result2}(b).
\\

\begin{algorithm}
\caption{Tournament selection for local optimized spin configuration}
\begin{algorithmic}
\State Choose the tournament size $K$
\State Choose the parent group set size $O$

\For {$i = 1$ to $O$}
\State Choose $K$ spin configurations from the parent generation at random
\State Calculate the energies of these $K$ spin configurations 
\State Choose the best spin configuration with the lowest energy from the tournament
\EndFor

\State Output selected $O$ spin configurations
\end{algorithmic}
\label{alg1}
\end{algorithm}

\begin{algorithm}
\begin{algorithmic}
\caption{Rank selection for local optimized spin configuration}
\State Choose the parent group set size $O$
\For {$i = 1$ to $O$}
\State Calculate the energies of all spin configurations from the parent generation
\State Choose and remove the best spin configuration with the lowest energy from the parent generation 

\EndFor

\State Output selected $O$ spin configurations
\label{alg2}

\end{algorithmic}
\end{algorithm}
\begin{algorithm}
\caption{Square crossover with 4 spin configurations}
\begin{algorithmic}
\State Choose 4 spin configurations in size of $N \times N$: S1,S2,S3,S4
\State Choose 2 random numbers $i$ and $j$ that allow $i+j=N$
\For { S1 to S4}
\State Split the spin configuration into 4 spin segments with the size of  $i \times i$, $i \times j$, $j \times i$, and $j \times j$  
\EndFor
\State Choose 4 different spin segments with size of $i \times i$, $i \times j$, $j \times i$, and $j \times j$ randomly from all spin segments.
\State Combine those 4 spin segments into one offspring configuration in size of $N \times N$

\State Output the offspring spin configuration
\end{algorithmic}
\label{alg3}
\end{algorithm}
\begin{algorithm}
\caption{Linear crossover with four spin configurations}
\begin{algorithmic}
\State Choose 4 spin configurations in size of $N \times N$: S1,S2,S3,S4
\State Choose 4 random numbers $i$, $j$, $k$, and $l$ that allow $i+j+k+l=N$
\For { S1 to S4}
\State Split the spin configuration into 4 spin segments with the size of  $i \times N$, $j \times N$, $k \times N$, and $l \times N$  
\EndFor
\State Choose 4 different spin segments with size of $i \times N$, $j \times N$, $k \times N$, and $l \times N$ randomly from all spin segments.
\State Combine those 4 spin segments into one offspring configuration in size of $N \times N$
\State Output the offspring spin configuration
\end{algorithmic}
\label{alg4}
\end{algorithm}

\subsection{Details of the simulated-annealing baseline simulation}
Simulated annealing was used as baseline in the analysis of our method, to facilitate comparison.
In these baseline simulations, we used a temperature mesh starting at 900~K and going down to to the target temperature. The temperature step size is adaptive and becomes increasingly dense with lower temperature. 
We used between 10 000 and 15 000 Metropolis steps at each temperature, with the larger number of steps for lower temperatures. The typical number of temperature steps were around 15. We also tried more dense temperature meshes, but they turned out to be significantly more time consuming and did not produce better results.

\begin{table*}
	\centering
	\caption{Relative energies of relevant different phases in the Pd/Fe/Ir (111) system (in mRy/atom) as a function of magnetic field $B$. FM stands for ferromagnetic, Sk stands for skyrmion. $6 \times 5$ SkL stands for a $6 \times 5$ skyrmion lattice, and $6 \times 6$ SkL stands for a $6 \times 5$ skyrmion lattice. The system size is $100\times100$ atomic spins}
	\label{tab:AverageRelativeEnergy}  
	\begin{tabular}{ccccc}
		\hline\hline\noalign{\smallskip}	
  $B$ (T)	& FM & Isolated Sk 	&  $6 \times 5$ SkL &	 $6 \times 6$ SkL \\
		\noalign{\smallskip}\hline\noalign{\smallskip}
  3.10&4.0845E$-$04&3.9266E$-$04&0.0000E$+$00&1.2199E$-$04\\
3.11&3.6309E$-$04&3.4878E$-$04  &0.0000E$+$00 &1.3504E$-$04\\
3.12&3.1772E$-$04&3.0488E$-$04	&0.0000E$+$00	&1.4808E$-$04\\
3.13&2.7236E$-$04	&2.6099E$-$04	&0.0000E$+$00	&1.6113E$-$04\\
3.14&2.2699E$-$04	&2.1710E$-$04	&0.0000E$+$00	&1.7417E$-$04\\
3.15&1.8163E$-$04	&1.7320E$-$04	&0.0000E$+$00	&1.8721E$-$04\\
3.16&1.3626E$-$04	&1.2931E$-$04	&0.0000E$+$00	&2.0025E$-$04\\
3.17&9.0900E$-$05	&8.5420E$-$05	&0.0000E$+$00	&2.1329E$-$04\\
3.18&4.5540E$-$05	&4.1530E$-$05	&0.0000E$+$00	&2.2634E$-$04\\
3.19&2.5300E$-$06	&0.0000E$+$00	&2.3700E$-$06	&2.4175E$-$04\\
3.20&1.0500E$-$06	&0.0000E$+$00	&4.6250E$-$05	&2.9868E$-$04\\
3.21&0.0000E$+$00	&4.1000E$-$07	&9.0560E$-$05	&3.5603E$-$04\\
3.22&0.0000E$+$00	&1.8900E$-$06	&1.3592E$-$04	&4.1444E$-$04\\
3.23&0.0000E$+$00	&3.3600E$-$06	&1.8129E$-$04	&4.7285E$-$04\\
3.24&0.0000E$+$00	&4.8400E$-$06	&2.2666E$-$04	&5.3126E$-$04\\
3.25&0.0000E$+$00	&6.3100E$-$06	&2.7202E$-$04	&5.8966E$-$04\\
3.26&0.0000E$+$00	&7.7900E$-$06	&3.1739E$-$04	&6.4808E$-$04\\
3.27&0.0000E$+$00	&9.2500E$-$06	&3.6275E$-$04	&7.0648E$-$04\\
3.28&0.0000E$+$00	&1.0730E$-$05	&4.0812E$-$04	&7.6489E$-$04\\
3.29&0.0000E$+$00	&1.2200E$-$05	&4.5348E$-$04	&8.2330E$-$04\\
3.30&0.0000E$+$00	&1.3670E$-$05	&4.9885E$-$04	&8.8171E$-$04\\
		\noalign{\smallskip}\hline
	\end{tabular}
\end{table*}
    
\subsection{Support movies}
The evolutionary process of predicting ground state of Pd/Fe/Ir(111) system at 0.1 mK with 0.7T, 2.7T and 3.7T are shown in support movie 1, 2 and 3, respectively
\end{document}